%% file: main.tex
\documentclass{article}
\usepackage{repsize}
\usepackage{graphicx,amssymb,amsmath}
\usepackage{acronym}
\usepackage{defs}
\usepackage{todonotes}

\title{Opposing Half Guards\footnote{B.~J.~N.~and C.~S.~were supported by grants 2018-04001 (Nya paradigmer f\"{o}r autonom obemannad flygledning) and 2021-03810 (Illuminate: bevisbart goda algoritmer f\"{o}r bevakningsproblem) from the Swedish Research Council (Vetenskapsr\r{a}det). E.~K.~ was supported by a sabbatical grant from the University of Wisconsin - Oshkosh.}}

\author{Erik Krohn\thanks{Department of Computer Science, University of Wisconsin - Oshkosh, Oshkosh, USA, {\tt krohne@uwosh.edu}}
        \and
        Bengt J.~Nilsson\thanks{Department of Computer Science and Media Technology, Malm\"o University, Sweden, {\tt  bengt.nilsson.TS@mau.se}}
        \and
        Christiane Schmidt\thanks{Department of Science and Technology, Link\"oping University, Sweden, {\tt  christiane.schmidt@liu.se}}
}
\date{}

\input{math}

\input{ines}

\begin{document}
\maketitle

\begin{abstract}
We study the art gallery problem for opposing half guards: guards that can either see to their left or to their right only. We present art gallery theorems, show that the location of half guards in 2-guardable polygons is not restricted to extensions, show that the problem is NP-hard in monotone polygons, and present approximation algorithms for spiral and staircase polygons.
\end{abstract}

\input{introduction}
\input{preliminaries}
\input{agt}

\input{2g.tex}
\input{np-hardness}
\input{spiral}
\input{staircase}


\bibliographystyle{sccaptitleplain}
{
\bibliography{lit}
}

\end{document}

%% file: math.tex
\long\gdef\remove#1{}

\newcommand{\VP}[1]{\mmath{{\bf V}\!(#1)}}
\renewcommand{\P}{\mmathp{{\bf P}}}

\def\comic#1#2#3{\parbox{#1}{\centering\includegraphics[width=#1]{#2}\\{\footnotesize #3}}}
\def\comicII#1#2{\parbox{#1}{\centering\includegraphics[width=#1]{#2}}}

\def\dash---{\kern.16667em---\penalty\exhyphenpenalty\hskip.16667em\relax}

\newcommand{\V}{\operatorname{\mathcal{V}}}

\acrodef{AGP}{Art Gallery Problem}
\acrodef{WRP}{Watchman Route Problem}
\acrodef{TGP}{Terrain Guarding Problem}
\acrodef{ATGP}{Altitude Terrain Guarding Problem}
\acrodef{AGPF}{Art Gallery Problem with Fading}

%% file: ines.tex
{\end{list}}

{\end{list}}

{\end{list}}

{\end{list}}

{\end{list}}

{\end{list}}

{\end{list}}

{\end{list}}

{\end{list}}

{\end{list}}

{\end{list}}

{\end{list}}


%% file: introduction.tex
\section{Introduction}\label{sec:intro}
The \ac{AGP}, based on a question by Victor Klee, is one of the classical problems in Computational Geometry. Klee asked for the minimum number of stationary guards with 360$^\circ$ vision that we need to place to achieve complete visibility coverage of a polygon \P. Such a guard $g\in\P$ can see a point $p\in\P$ iff $\overline{gp}$ is fully contained in~$\P$.
Typical results can be classified in two categories:
\begin{enumerate}
\item ``Art Gallery Theorems'': Worst-case, combinatorial bounds on the number of VPs that are sometimes necessary and always sufficient to cover a class of polygons---bounds on the maximum value of $G(\P)$ over all polygons of $n$ vertices, $g(n)$. 
\item Computational complexity and algorithmic results for 
the minimization of the number of star-shaped polygons (the {\it visibility polygons} (VPs) of guards) that cover a polygon---computation of~$G(\P)$.  
\end{enumerate}

Results on (1) are presented in, e.g.,~\cite{c-ctpg-74,f-spcwt-78,kkk-tgrfw-83,o-agta-87}. Results settling the computational complexity, (2), are given in, e.g.,~\cite{aam-agpec-21,e-irgph-98,ll-ccagp-86,o-agta-87,os-snpdp-83}, approximation algorithms are given in, e.g.,~\cite{bm-aaagp-17,kk-iagsg-11}.

\begin{figure}   
\centering
\hfill
\comic{.2\textwidth}{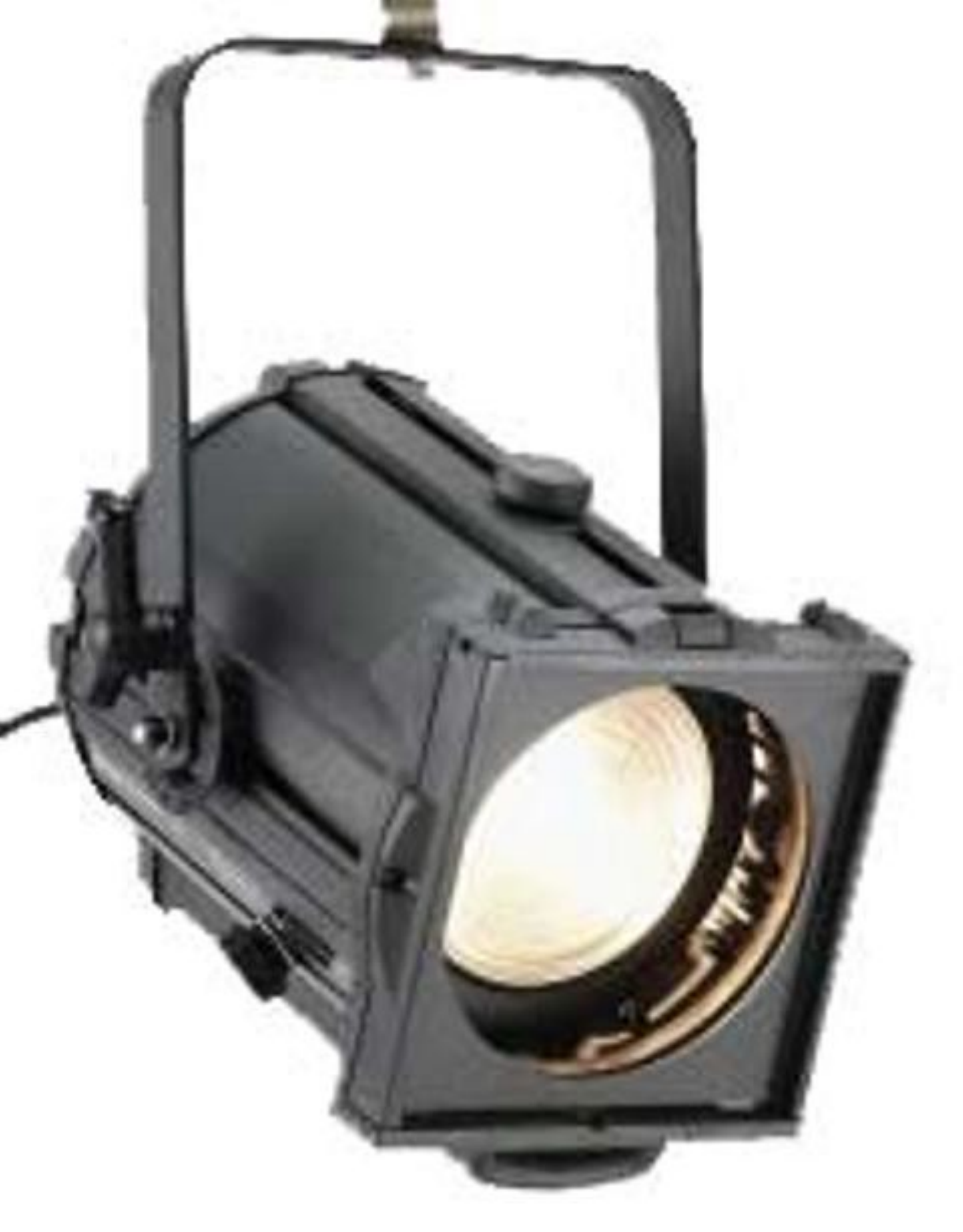}{(a)}\hfill\hfill
\comic{.4\textwidth}{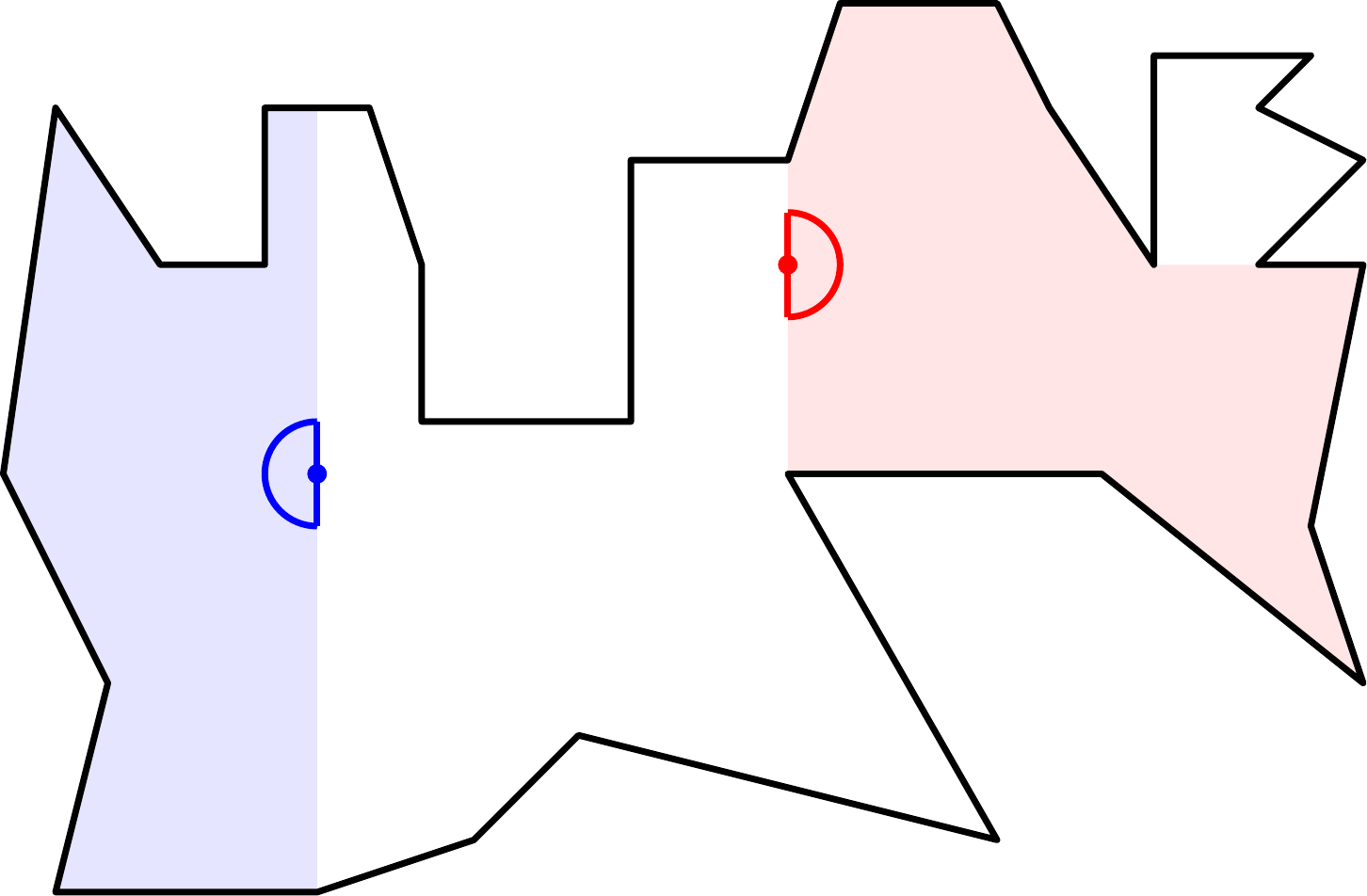}{(b)}
\hfill\mbox{}
  \caption{\label{fig:spotlight}(a) A spotlight (picture from flickr.com user Toby Simkin under licence CC BY-NC-SA 2.0.), (b) a polygon (black) with a left-looking half guard and its visibility polygon (blue) and a right-looking half guard and its visibility polygon (red).}
\end{figure}

Here, the guards do not have 360$^\circ$ vision, but every guard can either see to the left or the right: imagine a spotlight as shown in Figure~\ref{fig:spotlight}(a), for which the upper bow is fixedly mounted (and cannot rotate/yaw), so, the only degree of freedom for the spotlight is analogue to an aircraft pitch. If we fully utilize this movement, the spotlight illuminates one of the two halfplanes defined by the line that contains the upper bow. Formally, a left-looking (right-looking) half guard $g\in\P$ sees a point $p\in \P$ iff $\overline{gp}$ is fully contained in $\P$ and if $p$ does not have larger (smaller) $x$-coordinate than $g$, see Figure~\ref{fig:spotlight}(b) for an illustration. 
We call such guards {\em opposing\/} in contrast to half guards that can all only see in one direction. Half guards that all see in only one direction have been considered by Gibson et al.~\cite{gkr-gmphg-17}.

Of course, if we found a feasible solution for the \ac{AGP} with ``ordinary'' guards, placing a left- and a right-looking half guard for each guard yields a feasible solution also for the \ac{AGP} with half guards---usually this will not be optimal, for monotone mountains (where we can compute an optimal solution for the AGP in polynomial time~\cite{dfm-atggu-19}) it would directly yield a 2-approximation. In this paper, we study questions of both types (1) and (2) for half guards.

%% file: preliminaries.tex
\section{Notation and Preliminaries}\label{sec:not}
We let \P\ denote a simple polygon, $|\P|=n$. We let $r$ or $n_r$  denote the number of reflex vertices of $P$. A simple polygon $\P$ is $x$-{\it monotone} if the intersection $\ell\cap\P$ of $\P$ with any vertical line $\ell$ is a connected set. Any $x$-monotone polygon decomposes into two $x$-monotone polygonal chains between the rightmost and leftmost point of $\P$. An $x$-monotone polygon is a {\it monotone mountain} or {\it uni-monotone}, if one of its two chains ---w.l.o.g. the upper chain--- is a single horizontal segment, $\mathcal{H}$. A polygon $\P$ is {\it orthogonal} (or rectilinear) if all of its edges are axis-parallel, that is, either horizontal or vertical. An orthogonal polygon $\P$ is a {\it staircase polygon} if it is $x$- and $y$-monotone. We assume a leftmost (rightmost) point of a staircase polygon to also be its lowest (highest) point and denote this point by $p_{\ell}$ ($p_u$). We have two polygonal chains connecting $p_{\ell}$ and $p_u$, the chain for which there exists a point on the other chain with the same $x$-coordinate but larger $y$-coordinate is denoted as the lower chain, the other chain is the upper chain. 

Our half guards can either look to their left or their right, formally, we define left-looking and right-looking half guards. Let $p_x$ and $p_y$ define the $x$- and $y$-coordinate of a point $p$, respectively.
\begin{itemize}
\item A {\it left-looking half guard} $g\in\P$ can see a point $q\in\P$ iff ($\overline{gq}$ does not intersect $\P$'s boundary AND $g_x\geq p_x$); we say that $g$ {\it half sees} $q$.
\item A {\it right-looking half guard} $g\in\P$ can see a point $q\in\P$ iff ($\overline{gq}$ does not intersect $\P$'s boundary AND $g_x\leq p_x$); we say that $g$ {\it half sees} $q$.
\end{itemize}

Because we sometimes compare with ``normal'' visibility, we also define when a (full) guard $g\in\P$ {\it sees} a point $p\in\P$: when the line segment $\overline{gp}$ does not intersect $P$'s boundary. For a point $p$, we let  \VP{p}\ denote the half-visibility polygon of $p$ and $\V(p)$ denote the ``normal'' visibility polygon of $p$.
In a polygon $\P$, a set of witnesses $W$ is a set of points in $\P$, such that $\forall w_1, w_2\in W: \V(w_1)\cap\V(w_2)=\emptyset$.

%% file: agt.tex
\section{Art Gallery Theorems for Opposing Half Guards}\label{sec:agt}
In this section, we give Art Gallery Theorems, that is, statements of the type ``$x(n)$ guards are always sufficient and sometimes necessary for polygons with $n$ vertices'', for different polygon classes.

\begin{figure}   
\centering
\hfill
\comic{.25\textwidth}{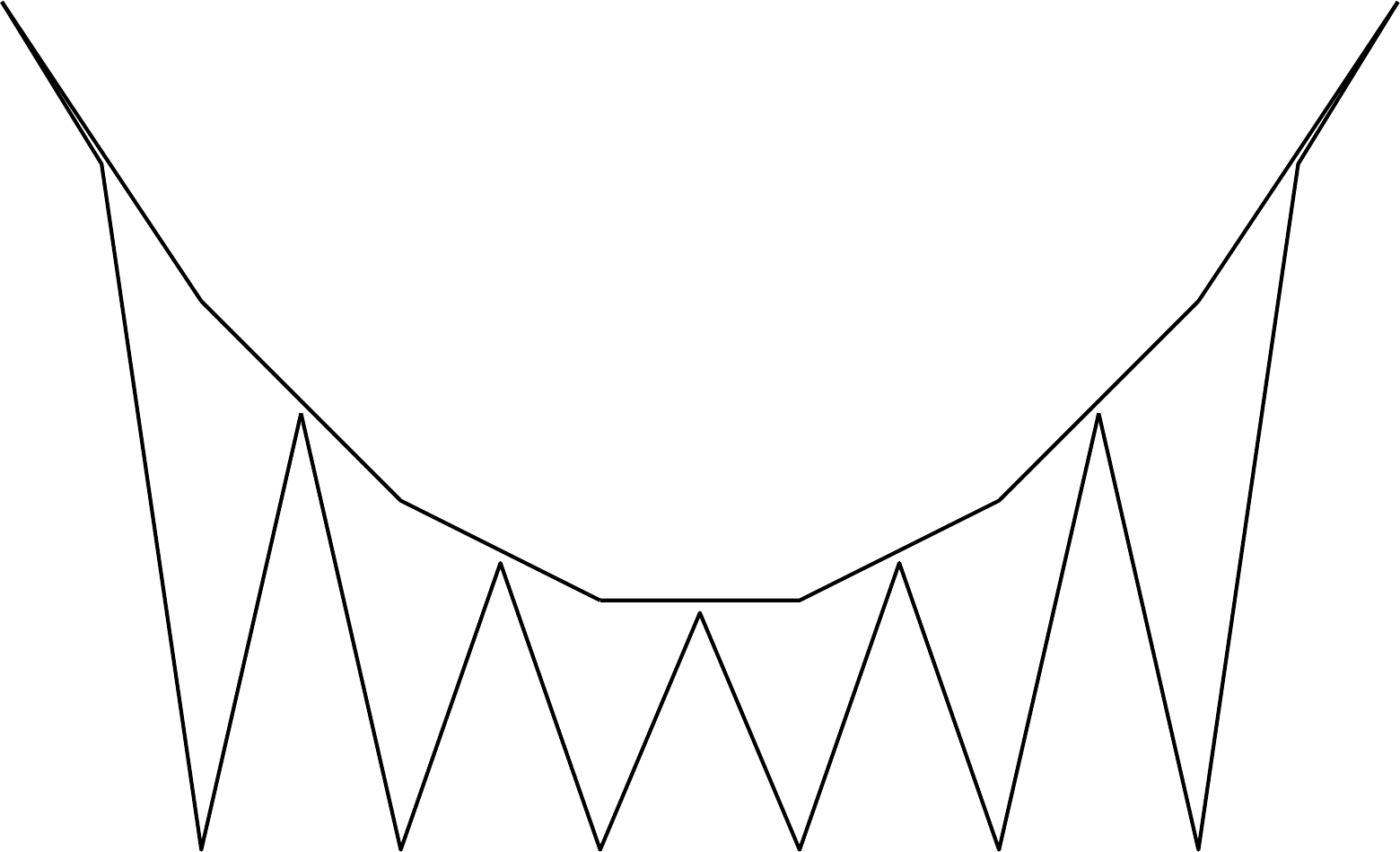}{(a)}\hfill
\comic{.20\textwidth}{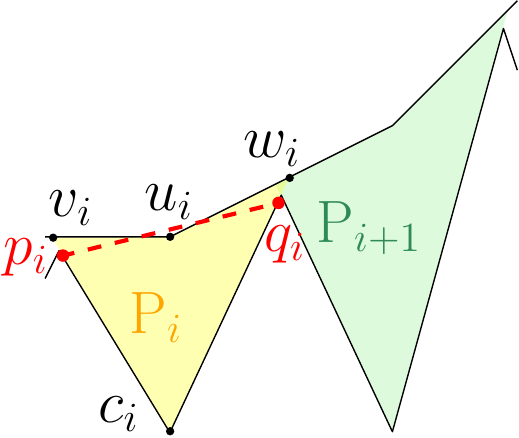}{(b)}\hfill
\comic{.25\textwidth}{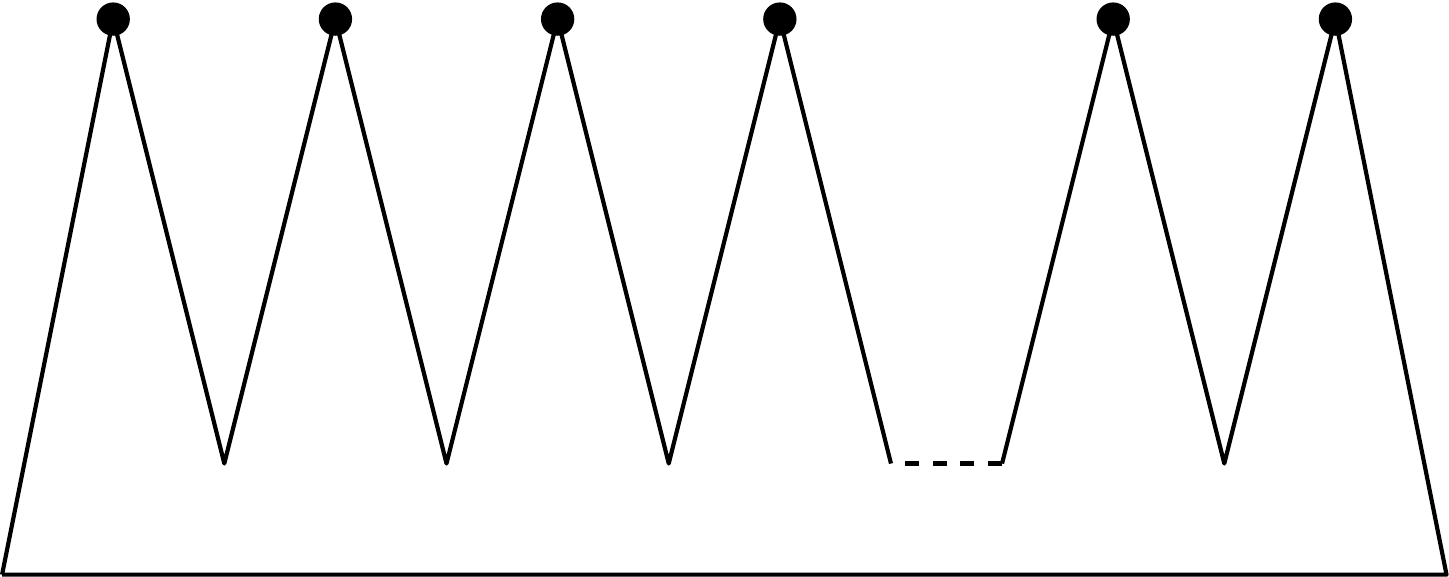}{(c)}
\hfill\mbox{}
 \caption{(a) Lower bound construction for simple (and monotone) polygons with $r>n/2$; (b) zoomed in on (a) showing $\mbox{\bf P}_i$ (yellow) and $\mbox{\bf P}_{i+1}$ (green). (c) Lower bound construction for simple (and monotone) polygons with $r\leq n/2$. }
  \label{fig:lb-mon}
\end{figure}

\begin{theorem}\label{th:agt-mon}
In simple polyons with $n$ vertices:
\begin{itemize}
\item For $r>n/2$: $2\lfloor n/3\rfloor$ half guards are always sufficient and sometimes necessary.
\item For $r\leq n/2$: $r + 1$ half guards are always sufficient and sometimes necessary.
\end{itemize}
\end{theorem}
\begin{proof}
For $r>n/2$, the upper bound follows trivially from the $\lfloor n/3\rfloor$ upper bound by Fisk~\cite{f-spcwt-78} for ``normal'' guards (triangulating the poylgon, three-coloring the vertices and using the least-frequently used color yields at most $\lfloor n/3\rfloor$): placing one right- and one left-looking half guard at each position of a ``normal'' guard results in $2\lfloor n/3\rfloor$ half guards.

For the lower bound, we construct a family of polygons $\P_{\!n}$, see Figure~\ref{fig:lb-mon}(a), that needs $2\lfloor n/3\rfloor$ half guards. 
$\P_{\!n}$ is an $x$-monotone polygon: the upper polygonal chain has reflex vertices only (except for the rightmost and leftmost vertex, which are convex), the lower chain has alternating reflex and convex vertices. The reflex vertices of the upper chain have the same $x$-coordinate as the convex vertices of the lower chain. The lower-chain vertices incident to the rightmost and leftmost vertex of $\P_{\!n}$ are either a reflex or convex vertex, such that we can define $\P_{\!n}$ also for $(n\bmod 3)\not\equiv 0$. For each convex vertex $c_i$ on the lower chain, we define a subpolygon $\P_i\subset\P_{\!n}$. We extend the two edges incident to $c_i$; let the two points where these extensions intersect with the upper chain be $v_i$ and $w_i$. Let the reflex vertex of the upper chain with the same $x$-coordinate as $c_i$ be $u_i$. The polygon $\P_i$ is defined by $c_i, v_i, u_i$ and $w_i$, see Figure~\ref{fig:lb-mon}(b). Note that $\P_i\cap\P_{i+1}\neq\emptyset$. We claim that we need two half guards per $\P_i$. 
Let $p_i$ and $q_i$ be two points on the edges incident to $c_i$ within distance $\varepsilon$ from the reflex vertices (marked in red in Figure~\ref{fig:lb-mon}(b)). The point $p_i$ can be seen from a right-looking half guard $g$ only if $g_x\leq p_{i_x}$, however---as indicated by the red line segment---such a half guard cannot see $q_i$. A similar argument holds for a left-looking half guard seeing $q_i$. Both $p_i$ and $q_i$ can be seen from points in $\P_i$ only. However, we saw that $\P_i$ and $\P_{i+1}$ overlap: assume that we place a left-looking half guard $g$ at $w_i$, it can see $q_i$ and $p_{i+1}$. We still need two (more) half guards in $\P_{i+1}$: a right-looking half guard at $w_i$ cannot see $q_{i+1}$, but a left-looking half guard that sees $w_i$ cannot see all of $\P_{i+1}$. Hence, we need $2\lceil n/3\rceil$ half guards.

For the upper bound for $r\leq n/2$, we recursively partition the polygon into $r+1$ convex pieces: we pick any reflex vertex and extend one of its incident edges until we hit the boundary. Then, in both subpolygons we created, this vertex is no longer a reflex vertex. Because we end up with convex pieces, we can cover each piece with either a left-looking half guard at its rightmost vertex or a right-looking guard at its leftmost vertex. This yields in total $r+1$ half guards.
For the lower bound, we construct a family of polygons $\P_{\!n}$, see Figure~\ref{fig:lb-mon}(c), that needs $r+1$ half guards. No two of the vertices marked with a point can be seen by a single half guard. Hence, we need $r+1$ half guards.
\end{proof}

The lower bound constructions for simple polygons are in fact both also monotone polygons, we yield: 
\begin{cortheorem}\label{cor:agt-mon}
In monotone polyons with $n$ vertices:
\begin{itemize}
\item For $r>n/2$: $2\lfloor n/3\rfloor$ half guards are always sufficient and sometimes necessary.
\item For $r\leq n/2$: $r + 1$ half guards are always sufficient and sometimes necessary.
\end{itemize}
\end{cortheorem}

\begin{figure}   
\centering
\hfill
\comic{.4\textwidth}{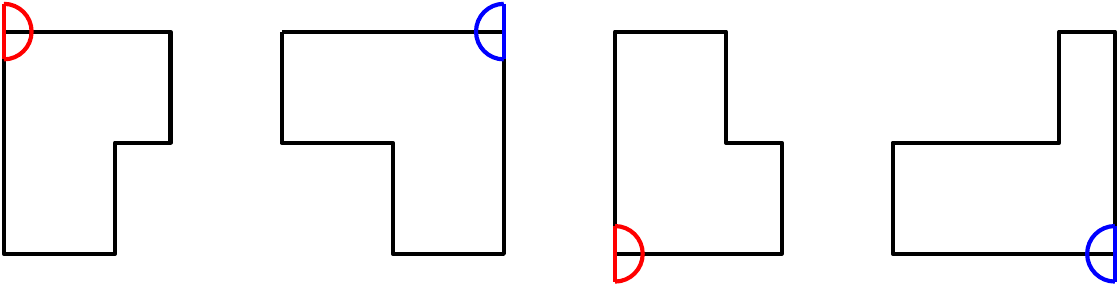}{(a)}\hfill\hfill
\comic{.42\textwidth}{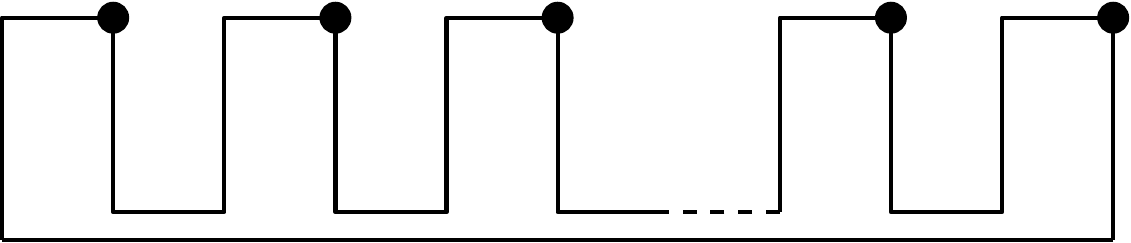}{(b)}
\hfill\mbox{}
  \caption{(a) Guarding L-shaped pieces with a single half guard; right-looking half guards are shown in red, left-looking half guards are shown in blue. (b) Lower bound construction for orthogonal polygons. }
  \label{fig:lb-orth}
\end{figure}

\begin{theorem}\label{th:agt-orth}
In simple orthogonal polyons with $n$ vertices, $\lfloor n/4\rfloor$ half guards are always sufficient and sometimes necessary.
\end{theorem}
\begin{proof}
An orthogonal polygon can be partitioned in $\lfloor n/4\rfloor$ L-shaped pieces (in linear time)~\cite{o-apragt-83,o-agta-87}. L-shaped pieces are orthogonal hexagons. Any L-shaped piece can be guarded by a single half guard placed at the only convex vertex, $v$, that can see all other vertices of the L-shaped piece (when the piece is considered as a simple polygon). Depending on whether the interior of the L-shaped piece lies in the left or right half plane of the vertical line through $v$, we use a left-looking or right-looking half guard, respectively. See Figure~\ref{fig:lb-orth}(a) for examples of guarding L-shaped pieces of the four possible orientations. This establishes that $\lfloor n/4\rfloor$ half guards are always sufficient.

Figure~\ref{fig:lb-orth}(b) shows a family of polygons $\P_{\!n}$ that needs $\lfloor n/4\rfloor$ half guards: no two of the vertices marked with a point can be seen by a single half guard. Hence, we need one half guard for every four edges.
\end{proof}

\begin{theorem}\label{th:agt-mon-mt}
In monotone mountains with $n$ vertices:
\begin{itemize}
\item For $r<n/2$: $r + 1$ half guards are always sufficient and sometimes necessary.
\item For $n/2\leq r\leq 3n/4$: $\lfloor n/2\rfloor$ half guards are always sufficient and sometimes necessary.
\item For $r>3n/4$: $2\cdot(n-r-2)\leq n/2$ half guards are sometimes necessary.
\end{itemize}
\end{theorem}
\begin{proof}
For $r\leq n/2$, the upper and lower bound follow as in the proof of Theorem~\ref{th:agt-mon} (i.e., the lower bound is shown in Figure~\ref{fig:lb-mon}(c) (where the lower horizontal chain is a horizontal segment)).

For the upper bound for $n/2\leq r\leq3n/4$, we consider the lower polygonal chain (the upper polygonal chain is a single horizontal segment $\mathcal{H}$). For guarding monotone mountains with ``normal'' guards, all guards can be placed on $\mathcal{H}$ and it is sufficient to guard all points of the lower polygonal chain to guard the complete polygon~\cite{dfm-atggu-19}, the arguments used there also hold for half guards. 

In between any pair of consecutive convex vertices, we have a reflex chain. One or two of the vertices in a reflex chain have a larger $y$-coordinate than the other vertices of that reflex chain. We split the reflex chain at one of these two vertices. Now, any convex vertex $v$ is adjacent to a split reflex chain both on its right, $C_R(v)$, and on its left, $C_{L}(v)$.  Let $\ell$ be the vertex with maximal $y$-coordinate in $C_L(v)$. We distinguish three cases:%
\begin{figure}   
\centering
\comicII{.8\textwidth}{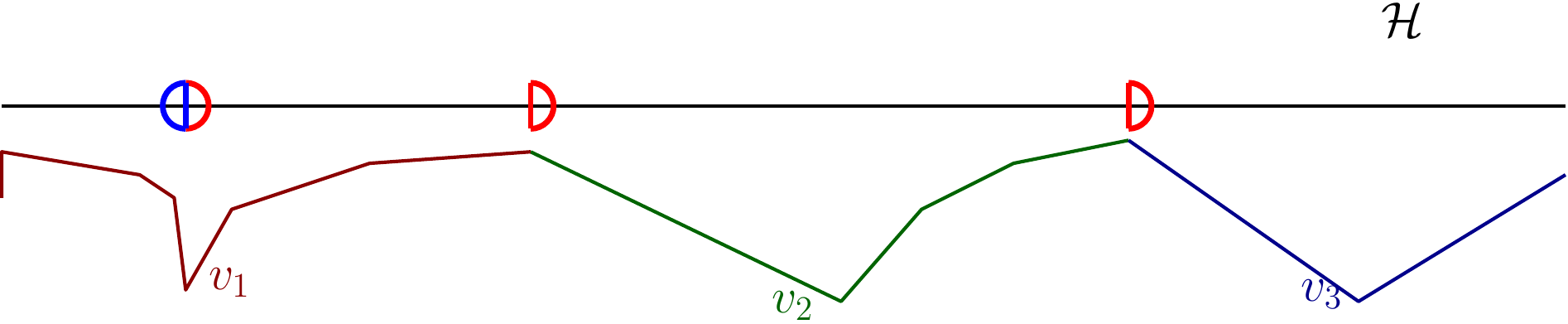}
  \caption{Upper bound construction for monotone mountains. The vertex $v_1$ and $C_R(v_1)$ and $C_L(v_1)$ are shown in dark red; $v_2$, $C_R(v_2)$, and $C_L(v_2)$ are shown in dark green; and $v_3$, $C_R(v_3)$, and $C_L(v_3)$ are shown in dark blue. Right-looking half guards are shown in red, left-looking half guards are shown in blue.}
  \label{fig:ub-mm-1}
\end{figure}
\begin{enumerate}
\item Both $C_R(v)$ and $C_L(v)$ have more than two edges: Let $h\in\mathcal{H}$ be the point with $h_x=v_x$, we place a right-looking and a left-looking guard at $h$. These half guards monitor $C_R(v)$ and $C_L(v)$, and we use two half guards for at least four edges. The vertex $v_1$ in Figure~\ref{fig:ub-mm-1} is an example for this case.
\item One of the two split reflex chains, w.l.o.g. $C_R(v)$, has more than two edges: let $h$ be the point on $\mathcal{H}$ with $h_x=\ell_x$. We place a right-looking half guard at $h$. This half guard sees $C_R(v)$ and $C_L(v)$ ($C_L(v)$ has only one edge), and we use one half guard for at least three edges. The vertex $v_2$ in Figure~\ref{fig:ub-mm-1} is an example for this case.
\item Both $C_R(v)$ and $C_L(v)$ have only one edge: let $h$ be the point on $\mathcal{H}$ with $h_x=\ell_x$. We place a right-looking half guard at $h$. It monitors $C_R(v)$ and $C_L(v)$, and we use one half guard for two edges. The vertex $v_3$ in Figure~\ref{fig:ub-mm-1} is an example for this case.
\end{enumerate}
In all cases, each half guard monitors on average at least two edges, and the claim follows.

For the lower bound for $n/2\leq r\leq3n/4$, we use a similar construction as in Figure~\ref{fig:lb-mon}(c), however, in between two consecutive convex vertices, we now include one, two or three reflex vertices, see Figure~\ref{fig:lb-mon-mt-1} for an example. If $C_R(v)$ and $C_L(v)$ have both one edge, we can guard both of these edges with a single (left- or right-looking) half guard.  If $C_R(v)$ and $C_L(v)$ have both two edges, these four edges cannot be seen by a single half guard, and no half guard from neighboring reflex chains can fully monitor these edges. Hence, we use one half guard per two edges.

\begin{figure}   
\centering
\comicII{.85\textwidth}{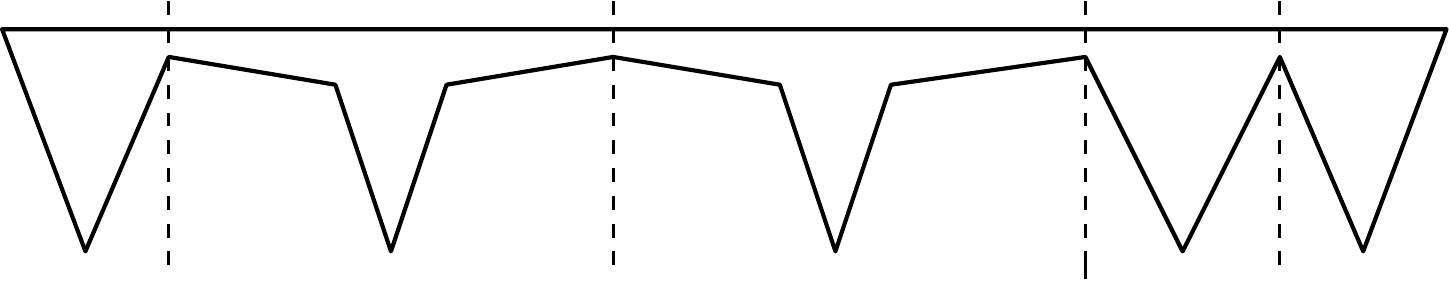}
  \caption{Lower bound construction for monotone mountains and $n/2\leq r\leq3n/4$.}
  \label{fig:lb-mon-mt-1}
\end{figure}

For $r>3n/4$, we only use case 1 (that is, only what is depicted in dark red in Figure~\ref{fig:ub-mm-1}): we place a right-looking and a left-looking guard on $\mathcal{H}$ at the $x$-coordinates of convex vertices---except for the leftmost and rightmost convex vertex that are located on $\mathcal{H}$. These guards see the complete lower polygonal chain. We have $c=n-r$ convex vertices, thus, we place $2(c-2)=2(n-r-2)$ half guards.
For the lower bound, we insert reflex chains as the dark-red reflex chains from Figure~\ref{fig:ub-mm-1} in the construction from Figure~\ref{fig:lb-mon-mt-1}.
\end{proof}

%% file: 2g.tex
\section{2-Guardable Polygons}\label{sec:2-g}

In his Master thesis, Belleville~\cite{b-ctcsp-91} showed 
that if a polygon $\P$ is two-guardable (two guards can fully monitor $\P$), $\P$ is two-guardable by two guards that are located on edge extensions (including the edges themselves). 
We show that this statement does not hold for half guards:

\begin{theorem}\label{th:2-g}
Let $\P$ be a polygon for which the minimum half-guard cover has cardinality two, let these two guards be denoted as $g_1$ and $g_2$. Then neither $g_1$ nor $g_2$ must be located on edge extensions. Moreover, neither $g_1$ nor $g_2$ must be located on polygon diagonals.
\end{theorem}
\begin{proof}
We construct a polygon $\P$ that can be covered by two half guards, but these guards do not lie on any edge extensions (or edges), that is, if we would restrict half-guard locations to edges and edge extensions, $\P$ cannot be covered by two half guards. 
The polygon $\P$ is shown in Figure~\ref{fig:2-g}. The shown half guards (one left-looking and one right-looking half guard) monitor $\P$, let the right-looking half guard be denoted as $g_r$ and the left-looking half guard be denoted as $g_{\ell}$. The lines of sight meeting in the niches on the top and bottom of $\P$ are shown in red and blue for $g_r$ and $g_{\ell}$, respectively. The dots in the polygon edges indicate very long edges.%
\begin{figure}   
\centering
\comicII{.85\textwidth}{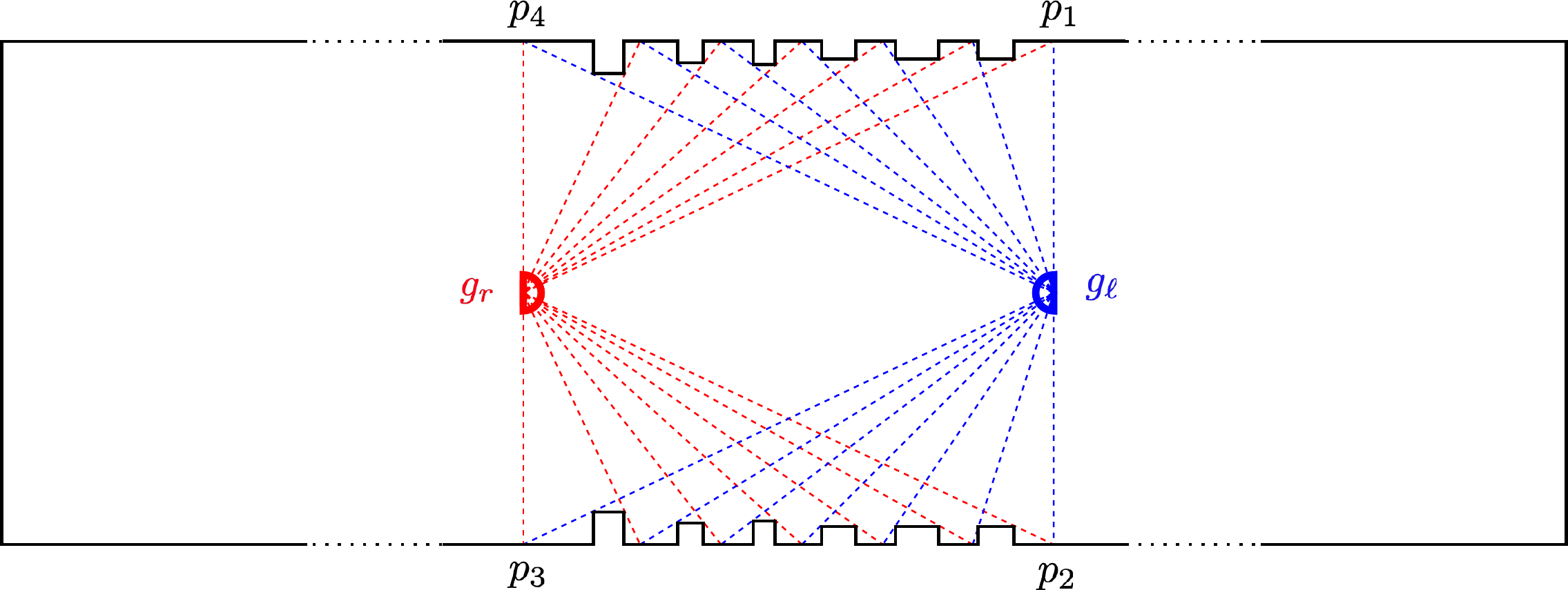}
  \caption{Polygon construction for Theorem~\ref{th:2-g}.}
  \label{fig:2-g}
\end{figure}

Clearly, as $\P$ is an orthogonal polygon, neither $g_r$ nor $g_{\ell}$ lie on any edge extensions, nor on any diagonals. 
It remains to show that no other pair of half guards monitors $\P$.

First, note that we cannot use two half guards that look into the same direction to monitor $\P$ with two half guards only: Then half guards cannot ``share'' seeing the niches, and we need one half guard per pair of mirrored niches. Hence, any minimum half-guard cover of $\P$ must contain a right-looking and a left-looking half guard.

Now assume that we try to move $g_r$ and $g_{\ell}$. Assume first that we only alter the $y$-coordinate of $g_r$. W.l.o.g.---the polygon is symmetric---we increase the $y$-coordinate of $g_r$ and obtain guard $g'_r$. Then, the first point that $g'_r$ sees on the top-right edge of $\P$ (the horizontal edge ending in the upper right vertex of $\P$) has a larger $x$-coordinate than 
$p_1$. Hence, we need to increase the $x$-coordinate of $g_{\ell}$. If we only alter the $x$-coordinate of $g_{\ell}$ to obtain $g'_{\ell}$, points at distance $\varepsilon$ from $p_3$ and $p_4$ on the same edges are not visible to $g'_{\ell}$, and we do no longer have a half-guard cover. Increasing also the $y$-coordinate of $g_{\ell}$ leaves part on the top unseen. Thus, assume that we increase the $x$-coordinate and decrease the $y$-coordinate of $g_{\ell}$. Then, a point at distance $\varepsilon$ from $p_3$ on its edge is not visible. 

Similar arguments yield that changing the $x$-coordinate of $g_r$ 
does not allow us to find a 
position for $g_{\ell}$ such that 
$\P$ is covered.
\end{proof}

Note that for ``normal'' guards, we may move the two guards to the $x$-coordinates of the left-most and right-most vertical edges of the niches.

%% file: np-hardness.tex
\section{Hardness Results for Opposing Half Guards}\label{sec:hard}
	NP-hardness for point guarding a monotone polygon with half guards that only see to the right was claimed in \cite{hkp-chgmp-22}. The same reduction can be used for opposing half guards. In \cite{hkp-chgmp-22}, the authors show an NP-hardness reduction from 3SAT. They show that certain vertices on the boundary represent truth values for variables in the original 3SAT instance. Clauses are represented by specific points on the boundary of the polygon. For example, if a clause $c = x_2 \vee x_5 \vee \overline{x_7}$ were in the original instance, then a vertex would exist on the boundary that would be seen by three vertices, namely the ones representing $x_2, x_5$ and $\overline{x_7}$. We briefly look at each pattern and show that if the polygon is guardable with $k$ guards, then all $k$ guards must be right-looking half guards.
	
\begin{figure}[ht]
    \centering
    \includegraphics[scale=0.46]{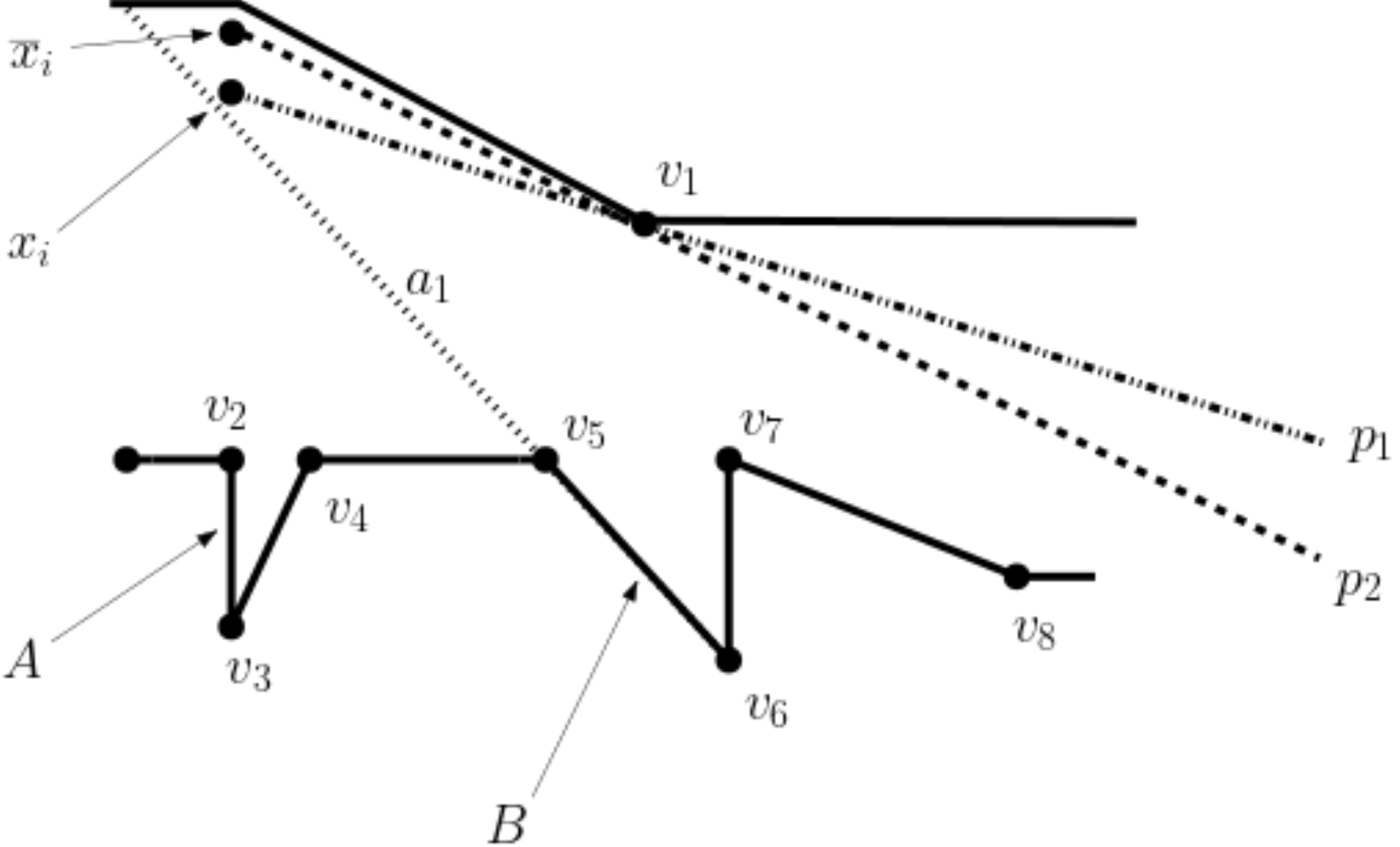}\hfill
    \includegraphics[scale=0.13]{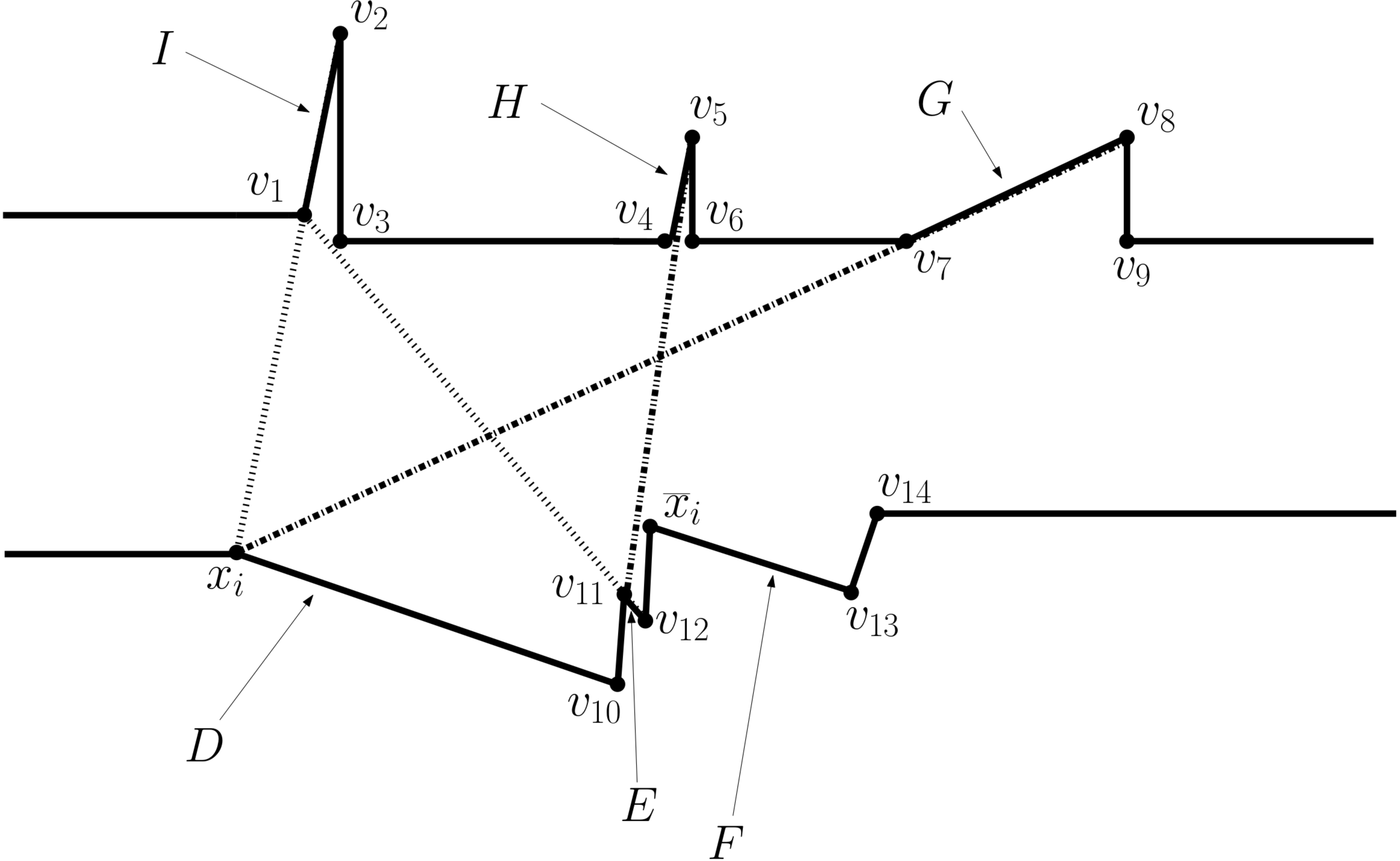}
    \caption{A starting pattern and a variable pattern for the NP-hardness reduction.}
    \label{fig:hardness}
\end{figure}
	
	\textbf{Starting Pattern:} No left-looking half guard can see both $v_3$ and $v_6$, see Figure \ref{fig:hardness}. No left-looking half guard placed outside of the starting pattern can see $v_3$ or $v_6$. Placing a left-looking half guard in the starting pattern for $x_i$ would require at least two guards to be placed for the $x_i$ starting pattern when one right-looking half guard is sufficient.

	\textbf{Variable Pattern:} Distinguished vertices of a variable pattern that can be seen by guards outside of the variable pattern are vertices $v_{10}$ and $v_{13}$, see Figure \ref{fig:hardness}. Neither of these vertices, and none of the other distinguished vertices in this pattern, can be seen by a left-looking half guard outside of the variable pattern. Any left-looking half guard placed inside the variable pattern will require too many guards to be placed to guard the entire variable pattern. In the original reduction, two right-looking guards are required to guard the variable pattern. Even with left-looking guards being allowed, two guards are still required.
	
	We note that only one of $v_{10}$ or $v_{13}$ will be seen by a guard to the left of the variable pattern. An incorrectly placed guard that sees both $v_{10}$ and $v_{13}$ will be an additional guard and will not reduce the number of guards needed in the current variable pattern. No guard (left or right-looking) can see both $v_2$ and $v_5$. Therefore, at least two guards are required to see $v_2$ and $v_5$ in each variable pattern. First, assume $v_{10}$ is seen by a right-looking half guard to the left of of the variable pattern. This leaves the following vertices to be guarded: $\{v_2, v_5, v_8, v_{12}, v_{13}\}$. No left-looking half guard can see more than one distinguished vertex in that list. If a left-looking half guard sees $v_2$, then the only location to see both $v_5$ and $v_{12}$ is a right-looking half guard at location $v_{11}$. If a left-looking guard is placed to see $v_5$, then the only location that sees $v_2$ and $v_{12}$ is a right-looking guard at location $v_1$. In both cases, neither $v_1$ nor $v_{11}$ sees $v_8$ and a third guard would be required.
	
	Next, assume $v_{13}$ is seen by a guard outside of the variable pattern. In this instance, no left-looking guard can see more than one of $\{v_2, v_5, v_8, v_{10}, v_{12}\}$. The same argument as above also applies here. Therefore, if a left-looking half guard is placed in this variable pattern, three guards are required when two right-looking half guards are sufficient. Those guard locations are $\{x_i, v_{11}\}$ or $\{\overline{x_i}, v_1\}$.
	
	No extra guards are required to see any of the clause distinguished points. Any left-looking half guard that is placed to see a clause distinguished point will only see that particular clause distinguished point. No other distinguished points in any starting or variable pattern will be seen by such a guard. If $k$ guards are sufficient to guard the entire polygon, then if a single left-looking half guard is placed in any pattern, an additional $k$ guards are required to see the entire polygon. Thus, the reduction from \cite{hkp-chgmp-22} holds even if left-looking half guards are allowed to be placed inside the monotone polygon.

%% file: spiral.tex
{
\input{mathspiral.tex}
\section{An Approximation Algorithm for Spiral Polygons}\label{sec:spiral}

\vspace*{-1ex}\noindent%
A simple polygon \P\ is {\em spiral\/} if it has two convex vertices \u\ and \u'\ such that a clockwise boundary walk from \u\ to \u'\ encounters only convex vertices and a counterclockwise boundary walk from \u\ to \u'\ encounters only reflex vertices. 
Nilsson and Wood~\cite{NilWood:spiral} show a linear time greedy algorithm to compute the minimum number of ``normal'' guards for spiral polygons. Replacing each full guard by two half guards at the same position and looking in opposing directions evidently gives a 2-approximation for opposing half guards. 
We show here a $3/2$-approximation for half guards based on dynamic programming.

For a half visibility polygon \VP{\p} of a point \p\ in polygon \P\!, we call a {\em window\/} a boundary edge of \VP{\p} that does not coincide with the boundary edges of~\P; see Figure~\ref{fig:spiral}.%
\begin{figure}
\begin{center}
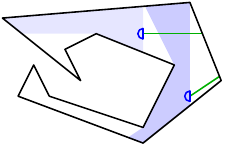
\end{center}
\caption{\label{fig:spiral}Windows of a spiral polygon. Interior half guards can always be moved to the boundary (along the green segment).}
\end{figure}%
\begin{lemma}\label{lem:window}
For a point \p\ in a spiral polygon \P\!, the half visibility polygon \VP{\p} has at most three windows.
\end{lemma}%
\begin{proof}
Consider any half-guard set \GG\ and pick a half-guard \g\ from \GG. If \g\ half-sees all of \P, then \g\ must by necessity lie on the convex boundary chain, otherwise \g\ has a $v$-window of non-zero length and the region behind the $v$-window is not seen.

If \g\ does not half-see any points of the reflex boundary chain, then \VP{\g} is a convex region and we can slide \g\ along its $v$-window until it reaches the boundary without losing coverage.

If \g\ half-sees points of the reflex boundary, consider the maximal subchain \CC{\g} of the reflex boundary chain that is half-seen by \g. If \CC{\g} is equal to the complete reflex boundary chain, then \g\ half-sees both \u\ and \u'\ and we can let \g\ slide along the extension of the segment $[\u,\g]$ away from \u\ until it hits the convex boundary without losing coverage. If \CC{\g} does not half-see the complete reflex boundary chain, then assume without loss of generality that \u\ is not half-seen by \g. If \u\ is seen, we use \u'\ instead in the following argument. Walk from \u\ along the reflex boundary chain until the first reflex vertex that is half-seen by \g. Denote this vertex \v. Vertex \v\ is an endpoint of \CC{\g} and we can let \g\ slide along the extension of the segment $[\v,\g]$ away from \v\ until it hits the convex boundary without losing coverage.

Thus, any half-guard set can be moved so that they all lie on the convex boundary chain while maintaining coverage and the size of the half-guard set.
\end{proof}
We denote the windows of \VP{\p} as the $v$-window, the vertical window through \p\, if it exists, and the at most two $r$-windows connecting a reflex vertex and the convex boundary chain of~\P; see Figure~\ref{fig:spiral}.

Next, we claim that we can assume that the half guards lie on the convex chain of~\P. 
\begin{lemma}\label{lem:convexboundary}
In a spiral polygon \P\!, there is an optimal set of half guards that all lie on the convex chain of\/~\P.
\end{lemma}
\begin{proof}
Consider any half-guard set \GG\ and pick a half-guard \g\ from \GG. If \g\ half-sees all of \P, then \g\ must by necessity lie on the convex boundary chain, otherwise \g\ has a $v$-window of non-zero length and the region behind the $v$-window is not seen.

If \g\ does not half-see any points of the reflex boundary chain, then \VP{\g} is a convex region and we can slide \g\ along its $v$-window until it reaches the boundary without losing coverage.

If \g\ half-sees points of the reflex boundary, consider the maximal subchain \CC{\g} of the reflex boundary chain that is half-seen by \g. If \CC{\g} is equal to the complete reflex boundary chain, then \g\ half-sees both \u\ and \u'\ and we can let \g\ slide along the extension of the segment $[\u,\g]$ away from \u\ until it hits the convex boundary without losing coverage. If \CC{\g} does not half-see the complete reflex boundary chain, then assume without loss of generality that \u\ is not half-seen by \g. If \u\ is seen, we use \u'\ instead in the following argument. Walk from \u\ along the reflex boundary chain until the first reflex vertex that is half-seen by \g. Denote this vertex \v. Vertex \v\ is an endpoint of \CC{\g} and we can let \g\ slide along the extension of the segment $[\v,\g]$ away from \v\ until it hits the convex boundary without losing coverage.

Thus, any half-guard set can be moved so that they all lie on the convex boundary chain while maintaining coverage and the size of the half-guard set.
\end{proof}

We let $n=n_c+n_r$ be the total number of vertices where $n_c$ is the number of convex vertices including \u\ and~\u'\ and $n_r$ is the number of reflex vertices. We order the vertices from \u\ to \u'\ so that $\u=\u_1,\ldots,u_{n_c}=\u'$ are the convex vertices and $v_1,\ldots,v_{n_r}$ are the reflex vertices in counterclockwise order starting from the vertex after \u\ and ending at the vertex before~\u'. To simplify, we let $v_0=\u$. We also denote by $e_i=[v_{i-1},v_i]$ the edge of the reflex boundary connecting $v_{i-1}$ and $v_{i}$, $1\leq i\leq n_r$.  

We identify special vertices that we call {\em corners}. A convex vertex $\u_j$ in \P\ is an {\em (outer) corner}, if the two incident edges to $\u_j$ both lie on the same side of a vertical line through $\u_j$ and assume for simplicity that \P\ has no vertical edges. We say that a corner is a {\em left corner\/} if the incident edges lie to the right of the vertical line through the corner, otherwise it is a {\em right corner}. Similarly, we can define the {\em inner corners\/} as the vertices of the reflex boundary chain for which the adjacent edges lie on the same side of the vertical line through the vertex.
As we follow the convex chain in clockwise order from a left corner to a right corner, we say that a left looking half guard is a {\em backward guard\/} and a right looking half guard is a {\em forward guard}. Similarly following the convex chain in clockwise order from a right corner to a left corner, a left looking half guard is a {\em forward guard\/} and a right looking half guard is a {\em backward guard}. 

In addition, we define some useful operations as follows:
\begin{itemize}
\item
For an edge $e$ of the reflex chain, the point \mext{e}\ is the point on the convex chain intersected by a ray exuded in the counterclockwise direction along~$e$. 

\item
For a point \p\ on the convex chain, the point \drp{\p}\ is the point on the reflex chain intersected by a ray exuded in the vertical direction from \p\ towards the interior of the polygon, if it exists, otherwise \drp{\p}\ is undefined.

\item
For a point \p\ on the convex chain, the point \spp{\p}\ is the point on the convex chain intersected by a ray exuded towards the last vertex on the reflex chain seen by a forward half guard at~\p.

\item
For a point \p\ on the reflex chain, the point \tp{\p}\ is the point on the convex chain intersected by a ray exuded in the vertical direction from \p\ towards the interior of the polygon. If \p\ is an inner corner, the ray is exuded in the vertical direction making \tp{\p}\ the furthest of the two possible points along the convex chain; see Figure~\ref{fig:ops}.

\item
Consider a set of half guards \GG\ on the convex chain and a half guard~\g\ placed at point \p\ of the convex chain. Assume that the half guards in \GG\ completely see the edges $e_1,\ldots,e_{i'-1}$ and \g\ completely sees the edges $e_{i'},\ldots,e_{i-1}$, for indices $i'\leq i$, then operation $\nxtG{\p}$ evaluates to the index $i$.  Our algorithm will always place guards so that they see a contiguous portion of the reflex chain starting at $v_0=\u$ and ending at some point \p\ on $e_i$, $i\geq1$. We can therefore assume that \GG\ is any set of half guards that see the edges before $e_{i'}$ and define $\nxt{\p}=i$ only based on the half guard~\g\ at~\p. %
\end{itemize}
\begin{figure}
\begin{center}
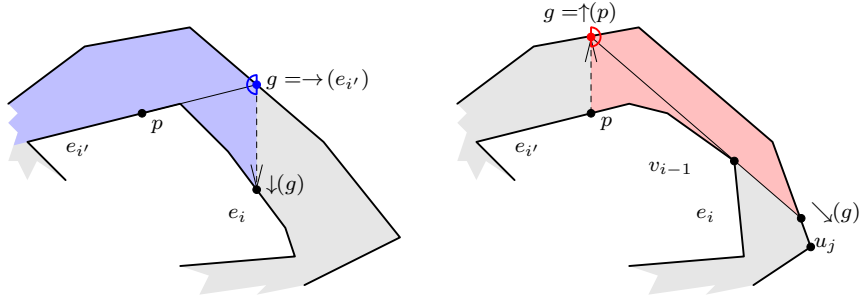
\end{center}
\caption{\label{fig:ops}Illustrating the four operations, a backward guard (left polygon) and a forward guard (right polygon)}
\end{figure}%

We specify our algorithm as a dynamic programming algorithm based on the following recursion. 
Let $i$ be the index of the furthest reflex edge $e_i$ not completely seen by the currently placed half guards, let \p\ be the last point on $e_i$ seen by the currently placed half guards, and let \q\ be the last point on the convex chain seen by the currently placed half guards.
The first call is~$G(0,\u,\u)$ and the cases are depicted in Figure~\ref{fig:spiralalgorithm}.

\begin{align*}
G\big(i,\p,\q\big)&=
\min\left\{
\begin{array}{lr}
G\big(\nxt{\u_j},v_{\nxt{\u_j}-1},\!\spp{\u_j}\big) + 1,    
& \mbox{half guard at next corner $\u_j$, if $\u_j$ sees \p\ and \q}
\\
G\big(\nxt{\g},\drp{\g},\g\big) + 1,                        
& \!\!\!\!\!\!\!\!\!\!\!\!\!\!\!\!\!\!\!\!\!\!\!\!\!\!\!\!\!\!\mbox{backward guard at $\g=\,\ext{e_i}$, if \g\ sees \q,}
\\ & \mbox{\drp{\g}\ is defined, and \g\ is before next corner}
\\
G\big(\nxt{\u_j},v_{\nxt{\u_j}-1},\!\spp{\u_j}\big) + 2,    
& \mbox{backward guard at $\g=\,\ext{e_i}$ and half guard}
\\
& \!\!\!\!\!\!\!\!\!\!\!\!\!\!\!\!\!\!\!\!\!\!\!\!\!\!\!\!\!\!\!\!\!\!\!\!\!\!\!\!\!\!\!\!\!\!\!\!\!\!\!\!\!\!\!\!\!\!\!\!\mbox{at next corner $u_j$, if \g\ sees \q, \drp{\g}\ is undefined, and \g\ is before $u_j$} 
\\
G\big(\nxt{\q},v_{\nxt{\q}-1},\!\spp{\q}\big) + 1,          
& \!\!\!\!\!\!\!\!\!\!\!\!\!\!\!\!\!\!\!\!\!\!\!\!\!\!\!\!\!\!\mbox{backward guard at $\g=\spp{\q}$ if $v_i$ does not see \q,}
\\
& \mbox{\drp{\g}\ is defined, and \g\ is before next corner} 
\\
G\big(\nxt{\u_j},v_{\nxt{\u_j}-1},\!\spp{\u_j}\big) + 2,    
& \mbox{backward guard at $\g=\spp{\q}$ and half guard}
\\
& \mbox{at next corner $u_j$, if $v_i$ does not see \q,} 
\\
& \mbox{\drp{\g}\ is undefined, and \g\ is before $u_j$} 
\\
G\big(\nxt{\q},v_{\nxt{\q}-1},\!\spp{\q}\big) + 1,          
& \!\!\!\!\!\!\!\!\!\!\!\!\!\!\!\!\!\!\!\!\!\!\!\!\!\!\!\!\!\!\mbox{forward guard at $\q$, if \q\ lies before \tp{\p}}
\\
& \mbox{on convex chain}  
\\
G\big(\nxt{\g},v_{\nxt{\g}-1},\!\spp{\g}\big) + 1,          
& \mbox{forward guard at $\g=\tp{p}$, if \tp{\p}\ lies before \q}
\\
& \mbox{on convex chain}  
\end{array}
\right.
\end{align*}
We ignore the bottom of the recursion, when $i>n_r+1$ as it follows the general description above without the recursive calls.
\begin{figure}
\begin{center}
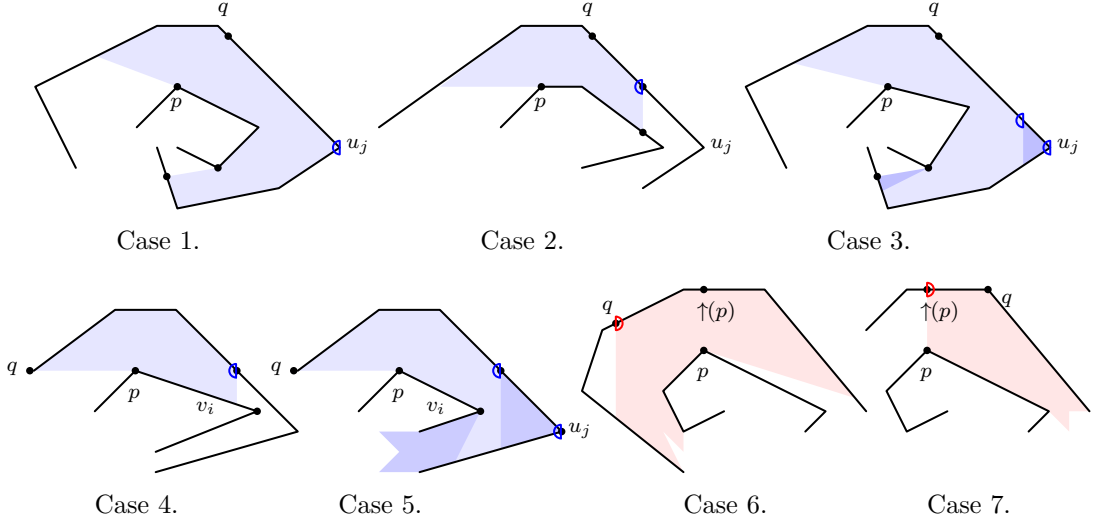
\end{center}
\caption{\label{fig:spiralalgorithm}Illustrating the seven cases in the recursion.}
\end{figure}

Each possible half guard position \q\ can be precomputed as the (outer) corners, the intersection points \mext{e}\ on the convex chain issuing from each edge $e$ of the reflex chain, and the intersection points \tp{v}\ on the convex chain issuing vertically from each vertex $v$ on the reflex chain, giving a linear number of possible positions. For each of these, we identify the linear sized sequence of continued intersection points \spp{\q}\ of the supporting segments with the convex boundary, giving at most a quadratic number of positions. Each position can be acquired in constant amortized time by a traversal of the two boundary chains taking quadratic time in total. 
The dynamic programming thus fills out a table of size $O(n)\times O(n^2)$, each position in constant time.

We next prove the correctness and approximation ratio of the algorithm.
\begin{lemma}\label{lem:threehalves}
If \OPT\!\ is a minimal set of half guards for a spiral polygon, then the algorithm covers the polygon and places at most $3|\OPT|/2+1$ half guards.
\end{lemma}
\begin{proof}
The correctness of the algorithm follows by construction, since it ensures that each window created by a half guard is seen by the next half guard placed.

It remains to show the approximation ratio for the algorithm. To do so, let \BB\ be the set of prespecified and computed points on the convex boundary chain of \P\ where the algorithm can place a half guard. We have $|\BB|\in O(n^2)$ from the discussion above. Consider an optimum set of half guards \OPT\ that we can assume by Lemma~\ref{lem:convexboundary} all lie on the convex chain. Follow the convex chain from \u\ to \u'\ until a half guard \ga\ in \OPT\ is reached that does not lie on a point in \BB. Let \p\ be the last point on the reflex chain seen by the guards in \OPT\ that lie before \ga\ in \OPT. If \ga\ is the first half guard, then $\p=\u$. We will show that we can exchange \ga\ and the subsequent half guard \gb\ in \OPT\ for three half guards \g, \g', and \g''\ that indeed lie on the corresponding points in \BB. Repeating the argument as the process follows the convex chain in clockwise order, proves our claim.

If \ga\ can be moved to the subsequent point in \BB, without losing visibility, we do so. This will not increase the number of guards.

Assume that \ga\ is a backward guard between two subsequent points \q\ and \q'\ in \BB\ and that \ga\ cannot be moved to \q'\ without losing visibility. Let $v_{i-1}$ be the first vertex on the reflex chain that \ga\ sees and thus \ga\ sees $e_i$. So, \p\ is some point on $e_i$ (except $v_{i}$). Since \ga\ cannot be moved to \q'\ and the reflex chain is seen by the previous guards until \p\ on $e_i$, there must be points on the convex chain that are not seen if \ga\ is moved towards \q'. Let \r\ be such a point on the convex chain and let \gz\ be the guard in \OPT\ that sees \r. The guard \gz\ must be a backward guard at \r\ and \r\ is by assumption in \BB, hence, \ga\ lies on the intersection of the convex chain and the supporting line from \r\ through the reflex chain and this point lies in \BB, a contradiction; see Figure~\ref{fig:spiralproof}(a).%
\begin{figure}
\begin{center}
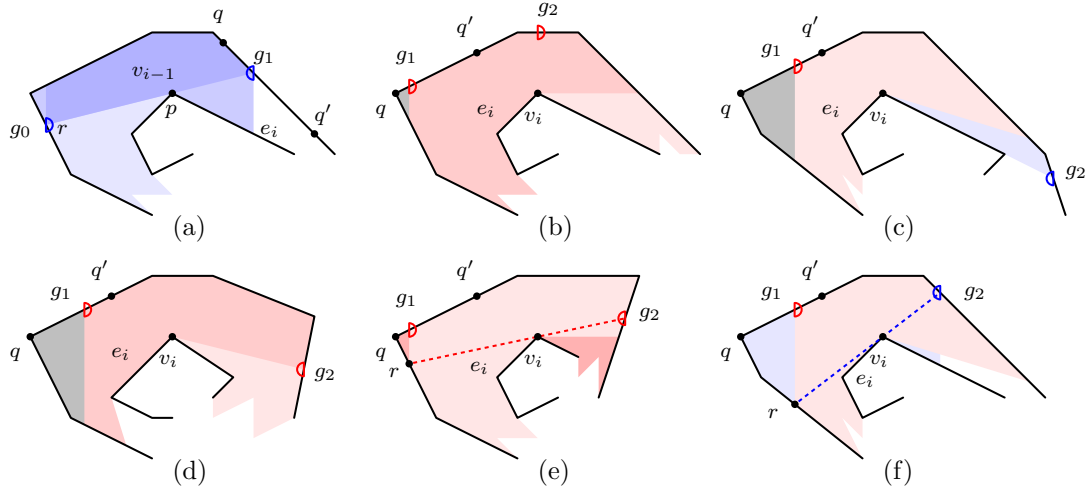
\end{center}
\caption{\label{fig:spiralproof}Illustrating the proof of Lemma~\protect\ref{lem:threehalves}. Grey regions are seen by previous half guards in~\protect\OPT\!.}
\end{figure}

Assume next that \ga\ is a forward guard between two subsequent points \q\ and \q'\ in \BB, that \ga\ cannot be moved to \q'\ without losing visibility, and that the region behind \ga\ is seen by the previous guards in \OPT\!\!. Let $v_i$ be the last vertex on the reflex chain that \ga\ sees and assume furthermore that \gb\ is a forward guard (or a backward guard with a corner between \ga\ and \gb\ along the convex chain).  Without loss of generality, we can assume that \gb\ lies on \spp{\ga}\ or on \tp{v_i}\ depending on whether \ga\ sees the next corner or not; see Figures~\ref{fig:spiralproof}(b),~(c), and~(d). If \gb\ does not lie on any of these points, we can move it there without losing visibility.
We can place two forward guards \g\ and \g'\ at \q\ and \q'. The half guard \g'\ sees a vertex $v_{i'}$ with $i\leq i'$, hence we can replace \gb\ by a half guard at \spp{\q'}, if \g'\ sees the next corner, or at \tp{v_{i'}}. Note that if \ga\ sees the next corner, so must \g'\!. In both cases, \ga\ and \gb\ are replaced with three guards at positions in~\BB.

Finally, assume that \ga\ is a forward guard, that the subsequent guard \gb\ in \OPT\ is backward (or forward with a corner between them) and the region behind \ga\ is not seen by previous guards in \OPT\!. We can argue for \gb\ as we did in the first case to obtain a point \r\ on the convex chain not seen by \gb\ if \gb\ is moved forward; see Figure~\ref{fig:spiralproof}(e) and~(f). The forward guard \ga\ lies on the intersection between vertical line segment intersection through \r\ and the convex chain (the vertical segment spans between \r\ and \ga). Thus, the convex region defined by the convex chain from \r\ to \ga\ and limited by the vertical segment through these points is seen by \gb\ and it contains a corner.
We can replace \ga\ and \gb\ by a half guard \g\ at the corner, a forward guard \g'\ at \tp{v_i}, where $v_i$ is the last reflex vertex seen be \g, and either a backward guard at \mext{e_{i'}}, where $e_{i'}$ is the first unseen edge of the reflex boundary, or a half guard at the subsequent corner, whichever point comes first along a traversal of the convex chain; see Figures~\ref{fig:spiralproof}(e) and~(f).

In each case, a pair of half guards from the optimum is replaced by a triple of half guards at a subset of our designated positions. Hence, there exists a feasible placement of half guards having size at most $\lceil3|\OPT|/2\rceil$. Since the dynamic programming algorithm computes the minimum such placement, the lemma follows.
\end{proof}
%
We have the following result.
\begin{theorem}\label{thm:spiral}
The algorithm described computes a $3/2$-approximate set of opposing half guards for spiral polygons in $O(n^3)$ time.
\end{theorem}
}

%% file: mathspiral.tex
%
%
%
%

%
\long\gdef\remove#1{}

\newcommand{\nxt}[1]{\mmathp{{\rm ix}~\!\!(#1)}}
\newcommand{\nxtG}[1]{\mmathp{{\rm ix}_{\GG}~\!\!(#1)}}
\newcommand{\ext}[1]{\mmathp{\rightarrow~\!\!\!(#1)}}
\newcommand{\mext}[1]{\mbox{\mmathp{\rightarrow\!\!(#1)}}}
\newcommand{\tp}[1]{\mmathp{\uparrow~\!\!\!\!(#1)}}
\newcommand{\drp}[1]{\mmathp{\downarrow~\!\!\!\!(#1)}}
\newcommand{\spp}[1]{\mmathp{\searrow~\!\!\!\!(#1)}}

%

%
\newcommand{\OPT}{\mmath{\GG^*}}
\newcommand{\GGs}{\mmath{\bar{\GG}}}
\newcommand{\GG}{\mmathp{{\cal G}}}
\newcommand{\CV}[2]{\mmath{{\cal C}_{#1}(#2)}}
\newcommand{\CC}[1]{\mmath{{\cal C}(#1)}}
\newcommand{\Ss}{\mmathp{{\cal S}}}

\newcommand{\Ga}{\mmath{{\GG}_1}}
\newcommand{\Gi}{\mmath{{\GG}_i}}
\newcommand{\Gj}{\mmath{{\GG}_j}}
\newcommand{\Gk}{\mmath{{\GG}_k}}

\newcommand{\BB}{\mmathp{{\cal B}}}
\newcommand{\FF}{\mmathp{{\cal F}}}

%
\renewcommand{\u}{\mmathp{u}}
\newcommand{\g}{\mmathp{g}}
\newcommand{\p}{\mmathp{p}}
\renewcommand{\r}{\mmathp{r}}
\newcommand{\q}{\mmathp{q}}
\renewcommand{\v}{\mmathp{v}}

\newcommand{\qu}{\mmathp{\q_{\u}}}

\newcommand{\ua}{\mmathp{\u_1}}
\newcommand{\ub}{\mmathp{\u_2}}
\newcommand{\ui}{\mmathp{\u_i}}
\newcommand{\uk}{\mmathp{\u_k}}

\newcommand{\gz}{\mmath{\g_0}}
\newcommand{\ga}{\mmath{\g_1}}
\newcommand{\gb}{\mmath{\g_2}}

\newcommand{\pa}{\mmathp{\p_1}}
\newcommand{\pb}{\mmathp{\p_2}}

%% file: Figures/spiral.pdf_t
\begin{picture}(0,0)%
\includegraphics[scale=2.5]{Figures/spiral.pdf}%
\end{picture}%
\setlength{\unitlength}{6315sp}%
\begingroup\makeatletter\ifx\SetFigFont\undefined%
\gdef\SetFigFont#1#2#3#4#5{%
  \reset@font\fontsize{#1}{#2pt}%
  \fontfamily{#3}\fontseries{#4}\fontshape{#5}%
  \selectfont}%
\fi\endgroup%
\begin{picture}(1074,733)(214,-257)
\end{picture}%

%% file: Figures/ops.pdf_t
\begin{picture}(0,0)%
\includegraphics[scale=0.8]{Figures/ops.pdf}%
\end{picture}%
\setlength{\unitlength}{3158sp}%
\begingroup\makeatletter\ifx\SetFigFont\undefined%
\gdef\SetFigFont#1#2#3#4#5{%
  \reset@font\fontsize{#1}{#2pt}%
  \fontfamily{#3}\fontseries{#4}\fontshape{#5}%
  \selectfont}%
\fi\endgroup%
\begin{picture}(6362,2263)(54,-1862)
\put(2101,-211){\makebox(0,0)[lb]{\smash{{\SetFigFont{8}{7.2}{\rmdefault}{\mddefault}{\updefault}{\color[rgb]{0,0,0}$\g=\ext{e_{i'}}$}%
}}}}
\put(2101,-1036){\makebox(0,0)[lb]{\smash{{\SetFigFont{8}{7.2}{\rmdefault}{\mddefault}{\updefault}{\color[rgb]{0,0,0}\drp{\g}}%
}}}}
\put(526,-736){\makebox(0,0)[lb]{\smash{{\SetFigFont{8}{7.2}{\rmdefault}{\mddefault}{\updefault}{\color[rgb]{0,0,0}$e_{i'}$}%
}}}}
\put(1201,-561){\makebox(0,0)[lb]{\smash{{\SetFigFont{8}{7.2}{\rmdefault}{\mddefault}{\updefault}{\color[rgb]{0,0,0}\p}%
}}}}
\put(1801,-1261){\makebox(0,0)[lb]{\smash{{\SetFigFont{8}{7.2}{\rmdefault}{\mddefault}{\updefault}{\color[rgb]{0,0,0}$e_{i}$}%
}}}}
\put(4726,-536){\makebox(0,0)[lb]{\smash{{\SetFigFont{8}{7.2}{\rmdefault}{\mddefault}{\updefault}{\color[rgb]{0,0,0}\p}%
}}}}
\put(5476,-1261){\makebox(0,0)[lb]{\smash{{\SetFigFont{8}{7.2}{\rmdefault}{\mddefault}{\updefault}{\color[rgb]{0,0,0}$e_{i}$}%
}}}}
\put(4276,314){\makebox(0,0)[lb]{\smash{{\SetFigFont{8}{7.2}{\rmdefault}{\mddefault}{\updefault}{\color[rgb]{0,0,0}$\g=\tp{\p}$}%
}}}}
\put(5101,-886){\makebox(0,0)[lb]{\smash{{\SetFigFont{8}{7.2}{\rmdefault}{\mddefault}{\updefault}{\color[rgb]{0,0,0}$v_{i-1}$}%
}}}}
\put(6401,-1486){\makebox(0,0)[lb]{\smash{{\SetFigFont{8}{7.2}{\rmdefault}{\mddefault}{\updefault}{\color[rgb]{0,0,0}$\u_{j}$}%
}}}}
\put(6376,-1261){\makebox(0,0)[lb]{\smash{{\SetFigFont{8}{7.2}{\rmdefault}{\mddefault}{\updefault}{\color[rgb]{0,0,0}$\spp{\g}$}%
}}}}
\put(4051,-736){\makebox(0,0)[lb]{\smash{{\SetFigFont{8}{7.2}{\rmdefault}{\mddefault}{\updefault}{\color[rgb]{0,0,0}$e_{i'}$}%
}}}}
\end{picture}%

%% file: Figures/spiralalgorithm.pdf_t
\begin{picture}(0,0)%
\includegraphics[scale=1.7]{Figures/spiralalgorithm.pdf}%
\end{picture}%
\setlength{\unitlength}{6710sp}%
\begingroup\makeatletter\ifx\SetFigFont\undefined%
\gdef\SetFigFont#1#2#3#4#5{%
  \reset@font\fontsize{#1}{#2pt}%
  \fontfamily{#3}\fontseries{#4}\fontshape{#5}%
  \selectfont}%
\fi\endgroup%
\begin{picture}(4027,1952)(-39,-1426)
\put(3206, 89){\makebox(0,0)[lb]{\smash{{\SetFigFont{8}{7.2}{\rmdefault}{\mddefault}{\updefault}{\color[rgb]{0,0,0}\p}%
}}}}
\put(3381,439){\makebox(0,0)[lb]{\smash{{\SetFigFont{8}{7.2}{\rmdefault}{\mddefault}{\updefault}{\color[rgb]{0,0,0}\q}%
}}}}
\put(3856,-61){\makebox(0,0)[lb]{\smash{{\SetFigFont{8}{7.2}{\rmdefault}{\mddefault}{\updefault}{\color[rgb]{0,0,0}$u_j$}%
}}}}
\put(3006,-436){\makebox(0,0)[lb]{\smash{{\SetFigFont{10}{9.6}{\rmdefault}{\mddefault}{\updefault}{\color[rgb]{0,0,0}Case~3.}%
}}}}
\put(1926, 89){\makebox(0,0)[lb]{\smash{{\SetFigFont{8}{7.2}{\rmdefault}{\mddefault}{\updefault}{\color[rgb]{0,0,0}\p}%
}}}}
\put(2101,439){\makebox(0,0)[lb]{\smash{{\SetFigFont{8}{7.2}{\rmdefault}{\mddefault}{\updefault}{\color[rgb]{0,0,0}\q}%
}}}}
\put(2576,-61){\makebox(0,0)[lb]{\smash{{\SetFigFont{8}{7.2}{\rmdefault}{\mddefault}{\updefault}{\color[rgb]{0,0,0}$u_j$}%
}}}}
\put(1726,-436){\makebox(0,0)[lb]{\smash{{\SetFigFont{10}{9.6}{\rmdefault}{\mddefault}{\updefault}{\color[rgb]{0,0,0}Case~2.}%
}}}}
\put(581, 89){\makebox(0,0)[lb]{\smash{{\SetFigFont{8}{7.2}{\rmdefault}{\mddefault}{\updefault}{\color[rgb]{0,0,0}\p}%
}}}}
\put(756,439){\makebox(0,0)[lb]{\smash{{\SetFigFont{8}{7.2}{\rmdefault}{\mddefault}{\updefault}{\color[rgb]{0,0,0}\q}%
}}}}
\put(1231,-61){\makebox(0,0)[lb]{\smash{{\SetFigFont{8}{7.2}{\rmdefault}{\mddefault}{\updefault}{\color[rgb]{0,0,0}$u_j$}%
}}}}
\put(381,-436){\makebox(0,0)[lb]{\smash{{\SetFigFont{10}{9.6}{\rmdefault}{\mddefault}{\updefault}{\color[rgb]{0,0,0}Case~1.}%
}}}}
\put(2051,-1111){\makebox(0,0)[lb]{\smash{{\SetFigFont{8}{7.2}{\rmdefault}{\mddefault}{\updefault}{\color[rgb]{0,0,0}$u_j$}%
}}}}
\put(1401,-986){\makebox(0,0)[lb]{\smash{{\SetFigFont{8}{7.2}{\rmdefault}{\mddefault}{\updefault}{\color[rgb]{0,0,0}\p}%
}}}}
\put(1201,-1411){\makebox(0,0)[lb]{\smash{{\SetFigFont{10}{9.6}{\rmdefault}{\mddefault}{\updefault}{\color[rgb]{0,0,0}Case~5.}%
}}}}
\put(951,-886){\makebox(0,0)[lb]{\smash{{\SetFigFont{8}{7.2}{\rmdefault}{\mddefault}{\updefault}{\color[rgb]{0,0,0}\q}%
}}}}
\put(1526,-1036){\makebox(0,0)[lb]{\smash{{\SetFigFont{8}{7.2}{\rmdefault}{\mddefault}{\updefault}{\color[rgb]{0,0,0}$v_i$}%
}}}}
\put(426,-986){\makebox(0,0)[lb]{\smash{{\SetFigFont{8}{7.2}{\rmdefault}{\mddefault}{\updefault}{\color[rgb]{0,0,0}\p}%
}}}}
\put(301,-1411){\makebox(0,0)[lb]{\smash{{\SetFigFont{10}{9.6}{\rmdefault}{\mddefault}{\updefault}{\color[rgb]{0,0,0}Case~4.}%
}}}}
\put(-24,-886){\makebox(0,0)[lb]{\smash{{\SetFigFont{8}{7.2}{\rmdefault}{\mddefault}{\updefault}{\color[rgb]{0,0,0}\q}%
}}}}
\put(676,-1036){\makebox(0,0)[lb]{\smash{{\SetFigFont{8}{7.2}{\rmdefault}{\mddefault}{\updefault}{\color[rgb]{0,0,0}$v_i$}%
}}}}
\put(2176,-661){\makebox(0,0)[lb]{\smash{{\SetFigFont{8}{7.2}{\rmdefault}{\mddefault}{\updefault}{\color[rgb]{0,0,0}\q}%
}}}}
\put(2526,-911){\makebox(0,0)[lb]{\smash{{\SetFigFont{8}{7.2}{\rmdefault}{\mddefault}{\updefault}{\color[rgb]{0,0,0}\p}%
}}}}
\put(2526,-686){\makebox(0,0)[lb]{\smash{{\SetFigFont{8}{7.2}{\rmdefault}{\mddefault}{\updefault}{\color[rgb]{0,0,0}\tp{\p}}%
}}}}
\put(2476,-1411){\makebox(0,0)[lb]{\smash{{\SetFigFont{10}{9.6}{\rmdefault}{\mddefault}{\updefault}{\color[rgb]{0,0,0}Case~6.}%
}}}}
\put(3351,-911){\makebox(0,0)[lb]{\smash{{\SetFigFont{8}{7.2}{\rmdefault}{\mddefault}{\updefault}{\color[rgb]{0,0,0}\p}%
}}}}
\put(3351,-686){\makebox(0,0)[lb]{\smash{{\SetFigFont{8}{7.2}{\rmdefault}{\mddefault}{\updefault}{\color[rgb]{0,0,0}\tp{\p}}%
}}}}
\put(3651,-636){\makebox(0,0)[lb]{\smash{{\SetFigFont{8}{7.2}{\rmdefault}{\mddefault}{\updefault}{\color[rgb]{0,0,0}\q}%
}}}}
\put(3376,-1411){\makebox(0,0)[lb]{\smash{{\SetFigFont{10}{9.6}{\rmdefault}{\mddefault}{\updefault}{\color[rgb]{0,0,0}Case~7.}%
}}}}
\end{picture}%

%% file: Figures/spiralproof.pdf_t
\begin{picture}(0,0)%
\includegraphics[scale=1.7]{Figures/spiralproof.pdf}%
\end{picture}%
\setlength{\unitlength}{6710sp}%
\begingroup\makeatletter\ifx\SetFigFont\undefined%
\gdef\SetFigFont#1#2#3#4#5{%
  \reset@font\fontsize{#1}{#2pt}%
  \fontfamily{#3}\fontseries{#4}\fontshape{#5}%
  \selectfont}%
\fi\endgroup%
\begin{picture}(3930,1863)(-14,-1312)
\put(576, 89){\makebox(0,0)[lb]{\smash{{\SetFigFont{8}{7.2}{\rmdefault}{\mddefault}{\updefault}{\color[rgb]{0,0,0}\p}%
}}}}
\put(1126, 64){\makebox(0,0)[lb]{\smash{{\SetFigFont{8}{7.2}{\rmdefault}{\mddefault}{\updefault}{\color[rgb]{0,0,0}\q'}%
}}}}
\put(901,289){\makebox(0,0)[lb]{\smash{{\SetFigFont{8}{7.2}{\rmdefault}{\mddefault}{\updefault}{\color[rgb]{0,0,0}\ga}%
}}}}
\put(751,439){\makebox(0,0)[lb]{\smash{{\SetFigFont{8}{7.2}{\rmdefault}{\mddefault}{\updefault}{\color[rgb]{0,0,0}\q}%
}}}}
\put(  1, 14){\makebox(0,0)[lb]{\smash{{\SetFigFont{8}{7.2}{\rmdefault}{\mddefault}{\updefault}{\color[rgb]{0,0,0}\gz}%
}}}}
\put(451,229){\makebox(0,0)[lb]{\smash{{\SetFigFont{8}{7.2}{\rmdefault}{\mddefault}{\updefault}{\color[rgb]{0,0,0}$v_{i-1}$}%
}}}}
\put(926, 14){\makebox(0,0)[lb]{\smash{{\SetFigFont{8}{7.2}{\rmdefault}{\mddefault}{\updefault}{\color[rgb]{0,0,0}$e_i$}%
}}}}
\put(176, 14){\makebox(0,0)[lb]{\smash{{\SetFigFont{8}{7.2}{\rmdefault}{\mddefault}{\updefault}{\color[rgb]{0,0,0}\r}%
}}}}
\put(601,-361){\makebox(0,0)[lb]{\smash{{\SetFigFont{10}{9.6}{\rmdefault}{\mddefault}{\updefault}{\color[rgb]{0,0,0}(a)}%
}}}}
\put(1651,389){\makebox(0,0)[lb]{\smash{{\SetFigFont{8}{7.2}{\rmdefault}{\mddefault}{\updefault}{\color[rgb]{0,0,0}\q'}%
}}}}
\put(1426,289){\makebox(0,0)[lb]{\smash{{\SetFigFont{8}{7.2}{\rmdefault}{\mddefault}{\updefault}{\color[rgb]{0,0,0}\ga}%
}}}}
\put(1951,464){\makebox(0,0)[lb]{\smash{{\SetFigFont{8}{7.2}{\rmdefault}{\mddefault}{\updefault}{\color[rgb]{0,0,0}\gb}%
}}}}
\put(1901, 64){\makebox(0,0)[lb]{\smash{{\SetFigFont{8}{7.2}{\rmdefault}{\mddefault}{\updefault}{\color[rgb]{0,0,0}$v_{i}$}%
}}}}
\put(1351, 89){\makebox(0,0)[lb]{\smash{{\SetFigFont{8}{7.2}{\rmdefault}{\mddefault}{\updefault}{\color[rgb]{0,0,0}\q}%
}}}}
\put(1726, 89){\makebox(0,0)[lb]{\smash{{\SetFigFont{8}{7.2}{\rmdefault}{\mddefault}{\updefault}{\color[rgb]{0,0,0}$e_i$}%
}}}}
\put(1951,-361){\makebox(0,0)[lb]{\smash{{\SetFigFont{10}{9.6}{\rmdefault}{\mddefault}{\updefault}{\color[rgb]{0,0,0}(b)}%
}}}}
\put(301,-511){\makebox(0,0)[lb]{\smash{{\SetFigFont{8}{7.2}{\rmdefault}{\mddefault}{\updefault}{\color[rgb]{0,0,0}\q'}%
}}}}
\put(551,-836){\makebox(0,0)[lb]{\smash{{\SetFigFont{8}{7.2}{\rmdefault}{\mddefault}{\updefault}{\color[rgb]{0,0,0}$v_{i}$}%
}}}}
\put(  1,-811){\makebox(0,0)[lb]{\smash{{\SetFigFont{8}{7.2}{\rmdefault}{\mddefault}{\updefault}{\color[rgb]{0,0,0}\q}%
}}}}
\put(151,-586){\makebox(0,0)[lb]{\smash{{\SetFigFont{8}{7.2}{\rmdefault}{\mddefault}{\updefault}{\color[rgb]{0,0,0}\ga}%
}}}}
\put(1126,-886){\makebox(0,0)[lb]{\smash{{\SetFigFont{8}{7.2}{\rmdefault}{\mddefault}{\updefault}{\color[rgb]{0,0,0}\gb}%
}}}}
\put(376,-811){\makebox(0,0)[lb]{\smash{{\SetFigFont{8}{7.2}{\rmdefault}{\mddefault}{\updefault}{\color[rgb]{0,0,0}$e_i$}%
}}}}
\put(601,-1261){\makebox(0,0)[lb]{\smash{{\SetFigFont{10}{9.6}{\rmdefault}{\mddefault}{\updefault}{\color[rgb]{0,0,0}(d)}%
}}}}
\put(2926,389){\makebox(0,0)[lb]{\smash{{\SetFigFont{8}{7.2}{\rmdefault}{\mddefault}{\updefault}{\color[rgb]{0,0,0}\q'}%
}}}}
\put(3176, 64){\makebox(0,0)[lb]{\smash{{\SetFigFont{8}{7.2}{\rmdefault}{\mddefault}{\updefault}{\color[rgb]{0,0,0}$v_{i}$}%
}}}}
\put(2626, 89){\makebox(0,0)[lb]{\smash{{\SetFigFont{8}{7.2}{\rmdefault}{\mddefault}{\updefault}{\color[rgb]{0,0,0}\q}%
}}}}
\put(3001, 89){\makebox(0,0)[lb]{\smash{{\SetFigFont{8}{7.2}{\rmdefault}{\mddefault}{\updefault}{\color[rgb]{0,0,0}$e_i$}%
}}}}
\put(2776,314){\makebox(0,0)[lb]{\smash{{\SetFigFont{8}{7.2}{\rmdefault}{\mddefault}{\updefault}{\color[rgb]{0,0,0}\ga}%
}}}}
\put(3901,-136){\makebox(0,0)[lb]{\smash{{\SetFigFont{8}{7.2}{\rmdefault}{\mddefault}{\updefault}{\color[rgb]{0,0,0}\gb}%
}}}}
\put(3226,-361){\makebox(0,0)[lb]{\smash{{\SetFigFont{10}{9.6}{\rmdefault}{\mddefault}{\updefault}{\color[rgb]{0,0,0}(c)}%
}}}}
\put(2926,-511){\makebox(0,0)[lb]{\smash{{\SetFigFont{8}{7.2}{\rmdefault}{\mddefault}{\updefault}{\color[rgb]{0,0,0}\q'}%
}}}}
\put(3176,-836){\makebox(0,0)[lb]{\smash{{\SetFigFont{8}{7.2}{\rmdefault}{\mddefault}{\updefault}{\color[rgb]{0,0,0}$v_{i}$}%
}}}}
\put(2626,-811){\makebox(0,0)[lb]{\smash{{\SetFigFont{8}{7.2}{\rmdefault}{\mddefault}{\updefault}{\color[rgb]{0,0,0}\q}%
}}}}
\put(3526,-586){\makebox(0,0)[lb]{\smash{{\SetFigFont{8}{7.2}{\rmdefault}{\mddefault}{\updefault}{\color[rgb]{0,0,0}\gb}%
}}}}
\put(2801,-1036){\makebox(0,0)[lb]{\smash{{\SetFigFont{8}{7.2}{\rmdefault}{\mddefault}{\updefault}{\color[rgb]{0,0,0}\r}%
}}}}
\put(2776,-586){\makebox(0,0)[lb]{\smash{{\SetFigFont{8}{7.2}{\rmdefault}{\mddefault}{\updefault}{\color[rgb]{0,0,0}\ga}%
}}}}
\put(3226,-1261){\makebox(0,0)[lb]{\smash{{\SetFigFont{10}{9.6}{\rmdefault}{\mddefault}{\updefault}{\color[rgb]{0,0,0}(f)}%
}}}}
\put(3126,-911){\makebox(0,0)[lb]{\smash{{\SetFigFont{8}{7.2}{\rmdefault}{\mddefault}{\updefault}{\color[rgb]{0,0,0}$e_i$}%
}}}}
\put(1651,-511){\makebox(0,0)[lb]{\smash{{\SetFigFont{8}{7.2}{\rmdefault}{\mddefault}{\updefault}{\color[rgb]{0,0,0}\q'}%
}}}}
\put(1426,-611){\makebox(0,0)[lb]{\smash{{\SetFigFont{8}{7.2}{\rmdefault}{\mddefault}{\updefault}{\color[rgb]{0,0,0}\ga}%
}}}}
\put(1901,-836){\makebox(0,0)[lb]{\smash{{\SetFigFont{8}{7.2}{\rmdefault}{\mddefault}{\updefault}{\color[rgb]{0,0,0}$v_{i}$}%
}}}}
\put(1351,-811){\makebox(0,0)[lb]{\smash{{\SetFigFont{8}{7.2}{\rmdefault}{\mddefault}{\updefault}{\color[rgb]{0,0,0}\q}%
}}}}
\put(1951,-1261){\makebox(0,0)[lb]{\smash{{\SetFigFont{10}{9.6}{\rmdefault}{\mddefault}{\updefault}{\color[rgb]{0,0,0}(e)}%
}}}}
\put(2326,-661){\makebox(0,0)[lb]{\smash{{\SetFigFont{8}{7.2}{\rmdefault}{\mddefault}{\updefault}{\color[rgb]{0,0,0}\gb}%
}}}}
\put(1401,-886){\makebox(0,0)[lb]{\smash{{\SetFigFont{8}{7.2}{\rmdefault}{\mddefault}{\updefault}{\color[rgb]{0,0,0}\r}%
}}}}
\put(1701,-861){\makebox(0,0)[lb]{\smash{{\SetFigFont{8}{7.2}{\rmdefault}{\mddefault}{\updefault}{\color[rgb]{0,0,0}$e_i$}%
}}}}
\end{picture}%

%% file: staircase.tex
\section{2-Approximation for Staircase Polygons}\label{sec:stair}

Gibson et al.~\cite{gknrz-ngsp-19} show that staircase polygons allow for a 2-approximation for ``normal'' guards. Our algorithm is inspired by their CCCG algorithm (where CCCG stands for canonical convex corner guard).
\begin{figure}   
\centering
\hfill
\comic{.20\textwidth}{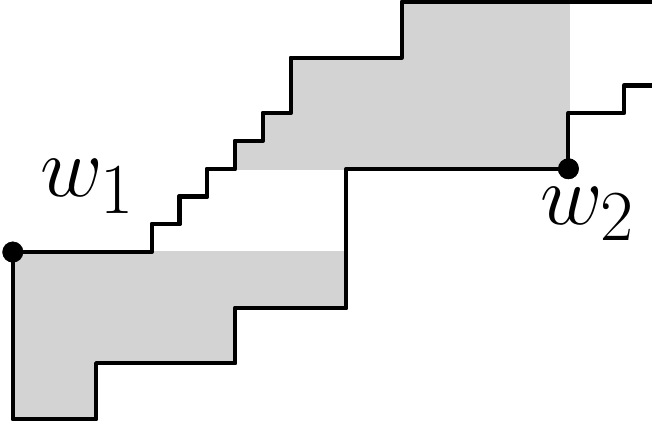}{(a)}\hfill
\comic{.20\textwidth}{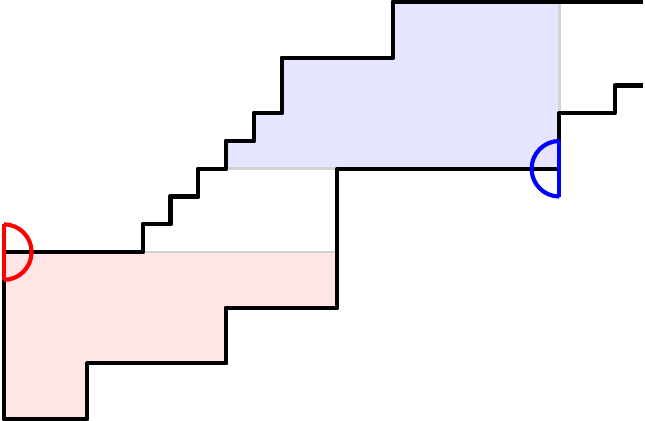}{(b)}\hfill
\comic{.20\textwidth}{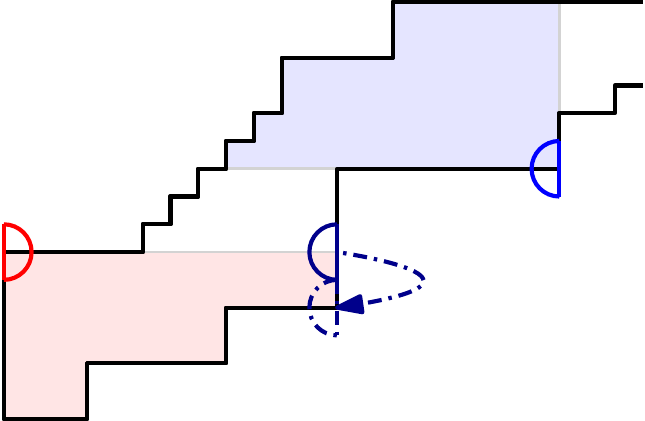}{(c)}
\hfill\mbox{}
  \caption{(a) A staircase polygon with CWS points $w_1$ and $w_2$, visibility polygons are shown in gray. (b) Placement of guards in $G_{cw}$. (c) Placement of guard in $G_s$ (darkblue) and moving the guard to a feasible vertex (dash-dotted). }
  \label{fig:stair}
\end{figure}

Let $G^*$ be an optimal opposing-half-guard set for a staircase polygon. We place a guard set $G$ that is composed of two sets, that is, we have $G=G_{cw}\cup G_s$. 

To construct $G_{cw}$, we place a set of witnesses on convex vertices, the {\it convex witness set} (CWS): We place witnesses alternatingly on (some) convex vertices of the lower and the upper chain. We place the first witness, $w_1$, on the first convex vertex of the upper chain that does not lie on $P$'s lowest edge. We then define $\P_i=\P\setminus \V(w_{i-1})$, where $\V(p)$ denotes the visibility polygon of $p$ under ``normal'' visibility. In $\P_i$ we place a witness on the first convex on the lower chain vertex that does not lie on $\P_i$s lowest edge, see Figure~\ref{fig:stair}(a) for an examplary witness placement. 
We yield 
$W=\{w_1, w_2,\ldots\}$. 
\begin{lemma}\label{le:cws}
$W$ is a set of witness points (and a CWS) and, hence, $|W|\leq |G^*|$.
\end{lemma}
\begin{proof}
The topmost edge of $\V(w_i)$, for $i=2k+1, k=0,\ldots$, is a horizontal edge, $e^h_i$, and $w_i$ and $e^h_i$ have the same $y$-coordinate. By construction, the lowest edge of $\V(w_{i+1})$, $e^h_{i+1}$, is also horizontal, and its $y$-coordinate is larger than that of $e^h_i$. An analogous argument holds for the vertical edges limiting $\V(w_i)$ and $\V(w_{i+1})$ for $i=2k, k=0,\ldots$. Thus, the visibility polygons of the points $w_i$ are pairwise disjoint. 
\end{proof}

We now place $G_{cw}$ as follows: place a right-looking half guard on each $w_i$ for which $i=2k+1, k=0,\ldots$, and a left-looking half guard on each $w_i$ for which $i=2k+1, k=1,\ldots$, see Figure~\ref{fig:stair}(b). Because the convex vertices are incident to vertical edges that limit their visibility to one half plane, the visibility polygons of the half guards coincide with the ``normal'' visibility polygons of our witnesses. We have $|G_{cw}|=|W|\leq |G^*|$.

For the construction of $G_s$, we consider the still unseen parts of $\P$: we can at most have $|G_{cw}|$ 
such polygon pieces (between each pair of witness visibility polygons and possibly one that includes either $\P$'s topmost or $\P$'s rightmost edge). We show that each such region is a staircase polygon for which either the upper or the lower chain has exactly two edges---a {\it stair}. Consider the placement of $w_{2k+1}$ and $w_{2(k+1)}$: We place $w_{2(k+1)}$ on the first convex lower chain vertex that does not have the same $y$-coordinate as $w_{2k+1}$. Hence, the lower chain of the polygonal region between $\VP{w_{2k+1}}$ and $\VP{w_{2(k+1)}}$ 
 consists of one horizontal edge (defined by the upper edge of $w_{2k+1}$'s visibility polygon) and one vertical edge (the upper end point of this edge has the same $y$-coordinate as $w_{2(k+1)}$).  Thus, the polygonal region is a stair. We can place a left-looking half guard at its lowest-rightmost point that covers it completely, see Figure~\ref{fig:stair}(c). Analgously, the polygonal region between $\VP{w_{2k}}$ and $\VP{w_{2k+1}}$ is a stair for which the upper chain has two edges, and we can guard it with a right-looking half guard at its highest-leftmost point. Thus, we have proved the following result.
\begin{theorem}\label{th:stair}
The set $G$ covers all of\/ $\P$ and 
$|G| = |G_{cw}| + |G_s| \leq 2\cdot|G_{cw}|  \leq 2\cdot|G^*| $. 
\end{theorem}

In fact, the result holds for vertex half guards: all half guards in $G_{cw}$ are already placed on vertices; we observe that we can slide each guard $g\in G_s$ between $w_{2k+1}$ and $w_{2(k+1)}$ down along the vertical edge it resides on without loosing coverage of its stair, see Figure~\ref{fig:stair}(c); analogously, a guard between $w_{2k}$ and $w_{2k+1}$ can be slided left along the horizontal edge it resides on. 